\documentclass[11pt]{article}
\usepackage{amsfonts}
\usepackage{amssymb}
\usepackage{amsmath}
\usepackage{latexsym}
\usepackage{graphicx}
\usepackage{mathrsfs}
\usepackage[english]{babel}
\usepackage[usenames]{color}
\input epsf

\renewcommand{\theequation}{\arabic{section}.\arabic{equation}}

\newcommand\be{\begin{equation}}
\newcommand\bea{\begin{eqnarray}}
\newcommand\ee{\end{equation}}
\newcommand\eea{\end{eqnarray}}
\newcommand\h{\frac{1}{2}}
\newcommand\Regge{\alpha'}
\newcommand{\bdm}{\begin{displaymath}}
\newcommand{\edm}{\end{displaymath}}

\newcommand{\mathsym}[1]{{}}

\newcommand{\f}[2]{\frac{#1}{#2}}

\newcommand{\p}{\phantom{a}}

\newcommand{\bref}[1]{(\ref{#1})}

\newcommand{\nn}{\nonumber \\}

\begin{document}

\begin{flushright}
\end{flushright}
\vspace{20mm}
\begin{center}
{\LARGE  Radiation from the non-extremal fuzzball}\\
\vspace{18mm}
{\bf Borun D. Chowdhury}$^{a,}$\footnote{E-mail: {\tt borundev@mps.ohio-state.edu}.}, 
{\bf and Samir D. Mathur}$^{a,}$\footnote{E-mail: {\tt mathur@mps.ohio-state.edu}.}\\
\vspace{8mm}
$^{a}$Department of Physics,\\ The Ohio State University,\\ Columbus,
Ohio, USA 43210\\
\vspace{4mm}
\end{center}
\vspace{10mm}
\thispagestyle{empty}
\begin{abstract}

The fuzzball proposal says that the information of the black hole state is distributed throughout the interior of the horizon in a `quantum fuzz'. There are special microstates where in the dual CFT we have `many excitations in the same state'; these are described by regular classical geometries without horizons. Jejjala et.al constructed non-extremal regular geometries of this type. Cardoso et. al then found that these geometries had a classical instability. In this  paper we show that the energy radiated through the unstable modes is exactly the Hawking radiation for these microstates. We do this by (i) starting with the semiclassical Hawking radiation rate (ii) using it to find the emission vertex in the CFT (iii) replacing the Boltzman distributions of the generic CFT state with the ones describing the microstate of interest (iv) observing that the emission now reproduces the classical instability.   Because the CFT has `many excitations in the same state' we get the physics of a Bose-Einstein condensate rather than a thermal gas, and the usually slow Hawking emission increases, by Bose enhancement, to a classically radiated field. This system therefore provides a complete gravity description of information-carrying radiation from a special microstate of the nonextremal hole.

\end{abstract}
\newpage
\setcounter{page}{1}

\section{Introduction}\label{intr}
\setcounter{equation}{0}

The fuzzball proposal makes a concrete statement about the structure of the black hole interior. The proposal says that the information of the hole is distributed throughout a horizon sized region. In the conventional picture of the black hole the region around the horizon is in the vacuum state;  Hawking radiation leaving from this horizon thus
carries no information of the hole. With the fuzzball proposal we do not have this problem: long range effects (identified in \cite{emission}) distribute the information of the hole all the way upto the horizon, and radiation leaving the fuzzball can carry information just like radiation from a piece of burning coal.

So far most of the work on constructing the fuzzball states has been in the context of extremal black holes. In the extremal case we have a well posed question: start with a microstate of string theory at weak coupling, and follow it to strong coupling. What does the microstate {\it look} like? The traditional picture of extremal holes
has an infinite length throat, which can however be traversed in finite proper time by an infalling observer. At the end of the throat is a horizon. Past this horizon is the `interior' of the hole and in this interior region is  a  singularity. In the fuzzball picture  the throat ends, after a long but finite distance, in a `fuzzy cap'. The `cap' is different for different microstates of the hole. To make explicit constructions, we can choose special microstates where in the dual CFT description we have a large number of excitations `in the same state'; this phrase will become more clear when we discuss the details of the CFT.  Such a choice of excitations gives the state a `classical limit', and the corresponding gravity solution is well-captured by supergravity, with a suppression of stringy corrections. This gravity solution is found to be regular, with no horizons, singularities or closed time-like curves. Such constructions for 2-charge extremal holes were made in \cite{lm4,lmm,st}, and for 3-charge holes in \cite{mss,gms1,gms2,lunin,bena,gimon,giustoRef}.\footnote{For reviews of such constructions see \cite{reviews}.}

There is a substantial body of literature now supporting the fact that all states of extremal holes are `fuzzball' states. What about non-extremal holes? Extremal holes have entropy like non-extremal holes, but they do not emit Hawking radiation. To fully establish the fuzzball proposal we would therefore like to study non-extremal states and see how they radiate.

In \cite{ross} {\it non-extremal} geometries were constructed, with  regularity properties  similar to those of the extremal geometries in \cite{gms1,gms2}. Should we think of these geometries as  particular  states of the nonextremal hole? Or are they  just solutions of Einstein's equations with no connection to black hole states? If the former is the case, then we should find that these states share some behavior with black hole states, after we take into account the somewhat special nature of our selected state arising from the fact that in the   dual CFT  we have  `many excitations in the same state'.

In \cite{myers,myersp} it was shown that the geometries of \cite{ross} have an `ergoregion instability' which causes them to decay. What is the relation of such  decay modes to the behavior of black holes? In the present paper, we show that these decay modes are {\it exactly} the `Hawking radiation' from these special microstates.

At first such a result may seem surprising, because the instability of \cite{myers} is a classical instability, while one normally thinks of Hawking radiation as a weak, quantum process. But we will see that this difference arises simply because in the  microstates under study  we have a `large number of CFT excitations in the same state'. Consider the radiation emitted by a gas of atoms. Each atom radiates independently, and if there are several different excited levels among the atoms then we will have a planckian distribution of radiated photons.  But now imagine that the atoms are all `in the same state' and when the radiate, they fall  into the {\it same} de-excited state. We then have a situation where the de-excited atoms are described by a Bose-Enstein condensate (BEC). After $n$ atoms have fallen to the lower energy state, the probability of de-exciting the next atom to this state is enhanced by  a factor $(n+1)$ compared to the probability of de-exciting the first atom. For large $n$, the radiation from the BEC appears like a classical, monochromatic wave, growing exponentially with time. 

 The same thing happens in the emission from our chosen black hole microstate, since in this special microstate we have `many excitations in the same state'. The same microscopic process that gives thermal-looking radiation from generic microstates gives monochromatic, exponentially growing emission from our special microstate. The frequency and emission rate of this radiation agrees exactly with the corresponding parameters of  the instability noted in \cite{myers}.

Let us recall some background facts and outline the details of our computation:

\bigskip

(a) We will consider  type IIB string theory, and compactify spacetime as $M_{9,1}\rightarrow M_{4,1}\times T^4\times S^1$. We have $n_1$ units of D1 charge along $S^1$, and $n_5$ units of D5 charge wrapped on $T^4\times S^1$. The bound state of the D1 and D5 charges can be described by an `orbifold CFT'. A  crude description of this CFT is in terms of an `effective string' which has total winding number $n_1n_5$ around the $S^1$. Massless excitations live on this effective string. These excitations move at the speed of light up and down the $S^1$ direction; these are called left and right moving excitations respectively. The degrees of freedom of the CFT arise from 4 bosons   $X^1, \dots X^4$, 4 real left moving fermions and 4 real right moving fermions.

(b)  We first look at the traditional semiclassical computation of Hawking radiation from the 4+1 dimensional non-extremal black hole carrying D1-D5 charges. Working at low energies, we obtain  the emission rate $\Gamma_l$ for different angular harmonics $l$ for a minimally coupled scalar.

(c) We recall the model for emission in the dual CFT \cite{dasmathur,mathurang,gubser}. Left moving and right moving excitations collide, and their energy is converted to gravitons $h_{ij}$ of the 10-dimensional string theory. Letting  the indices $i,j$ be along the $T^4$, we get emission of quanta that are minimally coupled scalars from the 4+1 dimensional point of view. The indices $i,j$ are carried by one left moving scalar $\partial X^i$ and one right moving scalar $\bar\partial X^j$. To get emission in the  angular harmonic $l$, we need $l$ left moving fermions and $l$ right moving fermions in addition to the scalars $X^i, X^j$.

In \cite{mathurang} the emission vertex coupling the CFT excitations to the outgoing scalars was constructed, but its overall normalization constant was not known. In the present paper we will call this normalization constant $V(l)$. To find the emission rate we have to multiply the vertex contribution by  the occupation numbers of the colliding modes; these are given by the thermal bose/fermi distributions $\rho_B(\omega), \rho_F(\omega)$.  It was shown in \cite{mathurang} that this CFT computation reproduces the  semiclassical Hawking radiation greybody factors,  upto the overall unknown constant $V(l)$. 

(d) By equating the semiclassical radiation rate to the CFT radiation rate we compute $
V(l)$. We could also have computed $V(l)$ directly from the orbifold CFT by using the techniques developed in \cite{lm1,lm2,lmpre}. In these references the CFT operators for insertion of angular momentum $l$ quanta was constructed, and 3-point functions computed. Recently \cite{recent} it has been shown that these 3-point functions agree exactly with 3-point functions in the dual $AdS$ geometry. But even though we could thus compute $V(l)$ from first principles, we will not do that. The reason is that our goal is to relate the instability of \cite{myers} to Hawking radiation, and the most direct  demonstration of this connection will be to get the $V(l)$ from Hawking radiation, make the changes needed to account for the special nature of our microstate, and show that we get the instability of \cite{myers}.

(e) Now consider the microstate geometries made in \cite{ross}. The corresponding CFT state is known through arguments similar to those used for extremal microstates in \cite{gms1,gms2}. The CFT states for our microstates of interest can be described as follows. The CFT state is in the completely `untwisted' sector; i.e., the effective string is made up of $n_1n_5$ separate `singly wound' loops, rather than one or more `multiwound loops'. Further, the only excitations on the loops are  fermions $\psi, \bar\psi$, and these fermions are in the lowest energy state for the spin they carry.  Each of the $n_1n_5$ loops is `in the same state'. So we see that the CFT states for the given geometry is rather special: in a generic state the loops would typically be twisted together to make longer loops, there would be both bosonic and fermionic excitations, the excitations need not be in the lowest possible energy state, and the spins of the fermions need not be aligned.

(f) Consider the computation of \cite{myers} The geometries of \cite{ross} were shown to have a classical instability, whose growth with time is described by a set of complex frequencies $\omega_k=\omega_k^R+i\omega_k^I$. This instability will cause the microstate geometry of \cite{ross} to emit a wave that will carry energy out to infinity.  We wish to derive this emission from a computation in the dual CFT. The emission in the CFT should consist of quanta that have energy equal to $\omega_k^R$, and we show that the spontaneous emission rate $\Gamma$  of these quanta should satisfy $\Gamma=2\omega^I_k$ (The factor of $2$ arises because $\omega_k^I$ gives the growth of the {\it amplitude} of the wave, while $\Gamma$ gives the {\it intensity}
of the radiation, a quantity that is quadratic in the amplitude.)

(g) With this background, we can describe the essence of the computation.  We consider the special CFT microstates that correspond to the geometries of \cite{ross}. We investigate which left and right moving excitations can collide to emit quanta with angular momentum $l$, and find the energy of the emitted scalar quanta from the energy of the CFT excitations participating in the interaction process. This energy agrees with the real part $\omega_k^R$ of the instability frequencies of \cite{myers}.  We then compute the rate $\Gamma$ of this emission in the CFT, using the normalization $V(l)$ of the interaction vertex found from the Hawking radiation process. We find exact agreement with the relation $\Gamma=2\omega_k^I$.

\bigskip

To summarize, we take the CFT computation of Hawking radiation, where the CFT excitations have thermal distributions  $\rho_B(\omega), \rho_F(\omega)$ for generic microstates. We replace these distributions by the actual occupation numbers for excitations in our specific microstate, and recompute the emission.  This emission is found to exactly match the emission of energy due to the classical instability. This result supports the picture that Hawking radiation is just the leakage of energy from a complicated `fuzzball cap';  simple caps like the ones in the geometries of \cite{myers} give rapid emissions of energy (manifested by classical instabilities), while the more complicated caps of generic microstates will give the slower emission corresponding to the Hawking emission rate.

\section{The non-extremal microstate geometries: Review}
\setcounter{equation}{0}

In this section we recall the microstate geometries that we wish to study, and explain how a suitable limit can be taken in which the physics can be described by a dual CFT.

\subsection{General nonextremal geometries}

Let us recall the setting for the geometries of \cite{ross}. Take type IIB string theory, and compactify 10-dimensional spacetime as
\be
M_{9,1}\rightarrow M_{4,1}\times T^4\times S^1
\ee
The volume of $T^4$ is $(2\pi)^4 V$ and the length of $S^1$ is $(2\pi) R$. The $T^4$ is described by coordinates $z_i$ and the $S^1$ by a coordinate $y$. The noncompact $M_{4,1}$ is described by a time coordinate $t$, a radial coordinate $r$, and  angular $S^3$  coordinates $\theta, \psi, \phi$. The solution will have angular momenta along $\psi, \phi$, captured by two parameters $a_1, a_2$.
The solutions will carry three kinds of charges. We have $n_1$ units of D1 charge along $S^1$, $n_5$ units of D5 charge wrapped on $T^4\times S^1$, and $n_p$ units of momentum charge (P) along $S^1$. 
These charges  will be described in the solution by three parameters $\delta_1, \delta_5, \delta_p$. We will use the abbreviations
\be
s_i=\sinh\delta_i, ~~~c_i=\cosh\delta_i, ~~~~(i=1, 5, p)
\ee
The metrics are in general non-extremal, so the mass of the system is more than the minimum needed to carry these charges. The non-extremality is captured by a mass parameter $M$. 

With these preliminaries, we can write down the solutions of interest. 
The general non-extremal 3-charge metrics with rotation were given in \cite{cy}
\begin{eqnarray} \label{3charge}
\mathrm{d}s^2&=&-\frac{f}{\sqrt{\tilde{H}_{1} \tilde{H}_{5}}}(
\mathrm{d}t^2 - \mathrm{d}y^2) +\frac{M}{\sqrt{\tilde{H}_{1}
\tilde{H}_{5}}} (s_p \mathrm{d}y - c_p
\mathrm{d}t)^2 \nonumber \\ &&+\sqrt{\tilde{H}_{1} \tilde{H}_{5}}
\left(\frac{ r^2 \mathrm{d}r^2}{ (r^2+a_{1}^2)(r^2+a_2^2) - Mr^2}
+\mathrm{d}\theta^2 \right)\nonumber \\ &&+\left( \sqrt{\tilde{H}_{1}
\tilde{H}_{5}} - (a_2^2-a_1^2) \frac{( \tilde{H}_{1} + \tilde{H}_{5}
-f) \cos^2\theta}{\sqrt{\tilde{H}_{1} \tilde{H}_{5}}} \right) \cos^2
\theta \mathrm{d} \psi^2 \nonumber \\ && +\left( \sqrt{\tilde{H}_{1}
\tilde{H}_{5}} + (a_2^2-a_1^2) \frac{(\tilde{H}_{1} + \tilde{H}_{5}
-f) \sin^2\theta}{\sqrt{\tilde{H}_{1} \tilde{H}_{5}}}\right) \sin^2
\theta \mathrm{d} \phi^2 \nonumber \\ && +
\frac{M}{\sqrt{\tilde{H}_{1} \tilde{H}_{5}}}(a_1 \cos^2 \theta
\mathrm{d} \psi + a_2 \sin^2 \theta \mathrm{d} \phi)^2 \nonumber \\ &&
+ \frac{2M \cos^2 \theta}{\sqrt{\tilde{H}_{1} \tilde{H}_{5}}}[(a_1
c_1 c_5 c_p -a_2 s_1 s_5 s_p) \mathrm{d}t + (a_2 s_1
s_5 c_p - a_1 c_1 c_5 s_p) \mathrm{d}y ] \mathrm{d}\psi \nonumber \\ 
&&+\frac{2M \sin^2 \theta}{\sqrt{\tilde{H}_{1} \tilde{H}_{5}}}[(a_2
c_1 c_5 c_p - a_1 s_1
s_5 s_p) \mathrm{d}t + (a_1
s_1 s_5 c_p - a_2 c_1 c_5 s_p) \mathrm{d}y] \mathrm{d}\phi \nonumber
\\ && + \sqrt{\frac{\tilde{H}_1}{\tilde{H}_5}}\sum_{i=1}^4
\mathrm{d}z_i^2
\end{eqnarray}
where
\begin{eqnarray} 
\tilde{H}_{i}=f+M\sinh^2\delta_i, \quad
f=r^2+a_1^2\sin^2\theta+a_2^2\cos^2\theta,
\end{eqnarray}
The D1 and D5 charges of the solution produce a RR 2-form gauge field given by  \cite{gms1} 
\begin{eqnarray} 
C_2 &=& \frac{M \cos^2 \theta}{\tilde H_1} \left[ (a_2 c_1
  s_5 c_p - a_1 s_1 c_5
  s_p) dt + (a_1 s_1 c_5 c_p - a_2 c_1 s_5 s_p) dy \right] \wedge d\psi
  \nonumber \\
&& + \frac{M \sin^2 \theta}{\tilde H_1} \left[  (a_1 c_1
  s_5 c_p - a_2 s_1 c_5
  s_p) dt  + (a_2 s_1 c_5 c_p - a_1 c_1 s_5 s_p) dy \right] \wedge d \phi
  \nonumber \\
&& - \frac{M s_1 c_1}{\tilde H_1} dt \wedge dy -
  \frac{M s_5 c_5}{\tilde H_1} (r^2 + a_2^2 + M
  s_1^2) \cos^2 \theta d\psi \wedge d\phi.
\end{eqnarray}
The angular momenta are given by
\bea
J_\psi &=& -  \f{\pi M}{4 G^{(5)}} (a_1 c_1 c_5  c_p - a_2 s_1 s_5 s_p)  \\
J_\phi &=& - \f{\pi M}{4 G^{(5)}} (a_2 c_1 c_5 c_p - a_1 s_1 s_5 s_p) 
\eea
and the mass is given by
\be
M_{ADM} = \f{\pi M}{4 G^{(5)}} (s_1^2 + s_5^2 + s_p^2 + \f{3}{2}) \label{Eqn:ADM_Mass}
\ee
It is convenient to define
\be 
Q_1=M\sinh\delta_1\cosh\delta_1, ~~Q_5=M\sinh\delta_5\cosh\delta_5, ~~Q_p=M\sinh\delta_p\cosh\delta_p
\label{qdef}
\ee
Extremal solutions are reached in the limit 
\be
M\rightarrow 0, ~~\delta_i\rightarrow\infty, ~~Q_i ~~{\rm fixed}
\ee
whereupon we get the BPS relation
\be
M_{extremal}={\pi\over 4 G^{(5)}} [Q_1+Q_5+Q_p]
\ee
The integer charges of the solution are related to the $Q_i$ through
\bea
Q_1&=& \frac{g \Regge^3}{V} n_1 \label{Eqn:Q1} \\ 
Q_5 &=& g \Regge n_5 \label{Eqn:Q5} \\
Q_p &=& \f{g^2 \Regge^4}{V R^2} n_p
\label{q1q5qp}
\eea

\subsection{Constructing regular microstate geometries}

The  solutions (\ref{3charge}) in general have  horizons and singularities.
One can take careful limits of the parameters in the
solution and find solutions which have {\it no} horizons or singularities.   In \cite{TwoChrSmooth}  regular 2-charge extremal geometries were found while in \cite{gms1,gms2} regular 3-charge extremal geometries were obtained. In \cite{ross} this method was used to obtain    regular 3-charge {\it non-extremal} geometries. These non-extremal geometries will be the ones of interest in the present paper.

The singularities at $\tilde{H}_i=0$ are curvature singularities while the relation $g(r)\equiv (r^2 + a_1^2)(r^2 + a_2^2) - Mr^2=0$ gives the positions of the horizons:
\be
r_\pm^2 = \h( M - a_1^2 -a_2^2) \pm \h \sqrt{ ( M - a_1^2 -a_2^2)^2- 4a_1^2 a_2^2}
\ee
In \cite{ross} smooth geometries were constructed by demanding that the singularity at $g(r_+)=0$ be a coordinate singularity similar to the singularity of polar coordinates at the origin of $\mathbb{R}^2$. With this condition we no longer have the region $r<r_+$ in our geometry. Since $r_-<r_+$,  the singularity  at $\tilde{H}_i=g(r_-)=0$ becomes irrelevant. 
In the general case $Q_p\ne 0$, there are four conditions for regularity \cite{ross}. (If $Q_p=0$ the conditions are somewhat simpler.) The first condition is
\be
M=a_1^2 + a_2^2 - a_1 a_2 \f{c_1^2 c_5^2 c_p^2 +s_1^2 s_5^2 s_p^2}{s_1 c_1 s_5 c_5 s_p c_p}
\ee
Two other conditions can be expressed by introducing the dimensionless parameters
\be
j \equiv \left(\f{a_2}{a_1} \right)^\h , \qquad s \equiv \left( \f{s_1 s_5 s_p}{c_1 c_5 c_p} \right)^\h \le 1 \label{Eqn:JSDef}
\ee
These two conditions are then
\be
\f{j+ j^{-1}}{s+s^{-1}} = m - n, \qquad \f{j - j^{-1}}{s- s^{-1}} = m+n, ~~~( m \in \mathbb{Z}, ~~n \in \mathbb{Z})\label{Eqn:3ChrSmoothCond1}
\ee
The fourth condition is
\be
R= \f{M s_1 c_1 s_5 c_5 ( s_1 c_1 s_5 c_5 s_p c_p)^\h}{\sqrt{a_1 a_2} ( c_1^2 c_5^2 c_p^2 - s_1^2 s_5^2 s_p^2)} \label{Eqn:3ChrSmoothCond2}
\ee
With these conditions we have some relations between the parameters of the geometry which will be of use later on:
\be
M = a_1 a_2 nm (s^{-2} -s^2)^2 \label{Eqn:SmoothMass}
\ee
\be
r_+^2 = -\f{M}{nm (s^{-2} -s^2)^2} s^2, \qquad r_-^2 =  -\f{M}{nm (s^{-2} -s^2)^2} s^{-2}  \label{Eqn:SmoothHorizon}
\ee
\bea
Q_p &=& nm \f{Q_1 Q_5}{R^2} ~~~~\Rightarrow ~~n_p = nm~ n_1 n_5 \\ \label{Eqn:ChargesInTermsOfCFT}
J_\psi &=& -m \f{\pi }{4 G^{(5)}} \f{Q_1 Q_5}{R} = -m ~n_1 n_5 \label{jpsival}\\
J_\phi &=& n \f{\pi }{4 G^{(5)}} \f{Q_1 Q_5}{R} = n~ n_1 n_5  \label{jphival} 
\eea
In the last three relations we have used (\ref{q1q5qp}) and 
\be
16 \pi G^{(10)} = (2 \pi)^5(RV) 16 \pi G^{(5)}=(2 \pi)^7 g^2 \Regge^4 \label{Eqn:NetwonsConstant}
\ee
The physical momentum along the $S^1$  will be
\be
P={n_p\over R}={nm~n_1n_5\over R}
\label{pvalue}
\ee

\subsection{The large $R$ limit for obtaining a  CFT dual}

A general geometry does not have a CFT dual. We can get a CFT dual if there is a region that behaves like an `asymptotically AdS' space to a good approximation. Starting at spatial infinity and  going inwards, the nature of our geometries is the following. There is flat space at infinity, then a `neck' region, then an AdS type geometry, then a `cap' which ends the space in a smooth way (with no horizon or singularity). We would like the AdS region to be `large', which means that its radial extent should be `many times the radius of curvature of the AdS'. As explained in \cite{what}, this situation is obtained if we take the limit where $Q_1, Q_5$ are fixed but\footnote{In the opposite limit $R \ll (Q_1Q_5)^{1\over 4}$ we get a thin black ring \cite{Giusto:2006zi}.}
\be
R\gg (Q_1Q_5)^{1\over 4}
\label{limit}
\ee
We will thus take the limit (\ref{limit}) in the solutions of \cite{ross}. In this limit it can be shown that
\be
M\rightarrow 0
\label{mmmm}
\ee
From (\ref{qdef}) we see that 
\be
\delta_1, \delta_5\rightarrow\infty
\ee
Thus we will have
\be
c_1 \simeq s_1, \qquad c_5 \simeq s_5 \label{Eqn:LargeRCoshSinhSame}
\ee
The geometry will have  a mass close to the mass of an extremal system carrying the D1 and D5 charges $n_1, n_5$; thus the dual CFT will describe low energy excitations of an `orbifold CFT' describing the bound state of the D1 and D5 charges. Let us compute the mass of the geometry above the mass of the extremal D1-D5 system.
From \bref{Eqn:3ChrSmoothCond1} we have
\be
j = m s- ns^{-1}, \qquad j^{-1} = ms^{-1} - n s
\ee
which gives
\be
s^2 + s^{-2} = \f{m^2 + n^2 -1}{mn} \label{Eqn:solS}
\ee
From \bref{Eqn:3ChrSmoothCond2} we get (using (\ref{Eqn:LargeRCoshSinhSame}))
\be
R \simeq \f{\sqrt{Q_1 Q_5}}{\sqrt{a_1 a_2}} \sqrt{s_p c_p}
\ee
which along with  \bref{Eqn:SmoothMass} gives
\be
M \simeq \f{Q_1 Q_5}{R^2} nm (s^{-2} - s^2)^2 s_p c_p
\ee
Using \bref{Eqn:LargeRCoshSinhSame} and \bref{Eqn:JSDef} we get
\be
s_p^2 \simeq \f{s^4}{1-s^4} 
\ee
which gives
\be
M \simeq \f{Q_1 Q_5}{R^2} nm (s^{-2} - s^2) \label{Eqn:MLargeR}
\ee
The mass of the extremal D1-D5 system is
\be
M_{extremal} = \f{\pi M}{4 G^{(5)}} (s_1^2 + s_5^2 + 1) \label{Eqn:extremal}
\ee
From \bref{Eqn:ADM_Mass} we see that the energy of the system above the energy of the extremal D1-D5 system is
\bea
\Delta M_{ADM} \simeq \f{\pi M}{8 G^{(5)}} (1+ 2 s_p^2) 
&\simeq& \f{\pi }{8 G^{(5)}}  \f{Q_1 Q_5}{R^2} nm (s^{-2} + s^2) \nonumber \\
&=&  \f{\pi }{8 G^{(5)}}  \f{Q_1 Q_5}{R^2} (m^2 + n^2 -1) \nonumber \\
&=& \f{1}{2 R} (m^2 + n^2 -1) n_1 n_5 
\label{massgeom}
\eea
where we used \bref{Eqn:solS},\bref{Eqn:Q1},\bref{Eqn:Q5} and \bref{Eqn:NetwonsConstant}. Note that this result is consistent with our initial observation (\ref{mmmm}) that $M$ becomes  small for large $R$. 

In the large $R$ limit that we have taken we also have, using   \bref{Eqn:SmoothHorizon} and \bref{Eqn:MLargeR}

\bea
r_+^2 &\approx& - \f{Q_1 Q_5}{R^2} \f{s^2}{s^{-2}  - s^2} \nonumber \\ 
r_-^2 &\approx& - \f{Q_1 Q_5}{R^2} \f{s^{-2}}{s^{-2}  - s^2} \label{Eqn:LargeRHorizon}
\eea
which gives
\be
r_+^2 - r_-^2 \approx  \f{Q_1 Q_5}{R^2} \label{Eqn:LargeRDiffHorizon}
\ee

\section{The instability of the geometries: Review}
\setcounter{equation}{0}

Shortly after the construction of the above 3-charge regular geometries it was shown in \cite{myers} that these geometries suffered from an instability. This was a classical ergoregion instability which is a generic feature of rotating non-extremal geometries. In this section we will reproduce the computations of \cite{myers} to find the complex eigenfrequencies for this instability in the large $R$ limit. 

\subsection{The wave equation for minimally coupled scalars}

We consider a minimally coupled scalar field in the 6-dimensional geometry obtained by dimensional reduction on the $T^4$.  Such a scalar arises for instance from $h_{ij}$, which is the graviton with both indices along the 
$T^4$. The wave equation for the scalar is
\be
\Box \Psi =0
\ee
We can separate variables with the ansatz \cite{larsen,ross,myers}\footnote{Our conventions are slightly different from those in \cite{myers}: we have the opposite sign of $\lambda$, for us positive $\omega$ will correspond to positive energy quanta, and for us $\omega$ has dimensions of inverse length.}
\be
\Psi = exp(- i \omega t + i \lambda \f{y}{R} + i m_\psi \psi + i m_\phi \phi) \chi(\theta) h(r)
\ee
which gives
\bea
\f{1}{\sin 2\theta} \f{d}{d \theta} \left( \sin 2\theta \f{d}{d \theta} \chi \right) + \left[\left(\omega^2-\f{\lambda^2}{R^2}\right) (a_1^2 \sin^2 \theta + a_2^2 \cos^2 \theta) - \f{m_\psi^2}{\cos^2 \theta}-\f{m_\phi^2}{\sin^2 \theta} \right] \chi + \Lambda \chi=0  \nonumber \\
\eea
\bea
\f{1}{r} \f{d}{dr} \left( \f{g(r)}{r} \f{d}{dr} h \right)+ (1- \nu^2) h -(r_+^2 - r_-^2) \left( \f{\zeta^2}{r^2 - r_+^2}  -\f{\xi^2}{r^2 - r_-^2}\right)h=0
\eea
where
\bea
g(r) &=& (r^2 - r_-^2)(r^2-r_+^2)  \nonumber \\
\xi &\equiv& \omega \varrho R - \lambda \vartheta - m_\phi n + m_\psi m \nonumber \\
\zeta &\equiv& -\lambda -m_\psi n + m_\phi m \nonumber \\
\varrho &\equiv& \f{c_1^2 c_5^2 c_p^2 - s_1^2 s_5^2 s_p^2}{s_1c_1 s_5 c_5} \nonumber \\
\vartheta &\equiv& \f{c_1^2 c_5^2 - s_1^2 s_5^2}{s_1 c_1 s_5 c_5} s_p c_p \nonumber \\
\nu^2 &\equiv& 1+ \Lambda - \left(\omega^2 - \f{\lambda^2}{R^2}\right)(r_+^2 + M s_1^2 +M s_5^2) - (\omega c_p -\f{\lambda}{R} s_p)^2M \label{Eqn:Definitions1}
\eea
Introducing the dimensionless radial coordinate
\be
x \equiv \f{r^2 - r_+^2}{r_+^2 - r_-^2}
\ee
 the radial equation becomes
\be
\partial_x[ x(x+1) \partial_x h] + \f{1}{4} \left[ \kappa^2 x + (1- \nu^2) + \f{\xi^2}{x+1} - \f{\zeta^2}{x} \right] h=0
\label{radial}
\ee
where
\be
\kappa^2 \equiv (\omega^2 - \f{\lambda^2}{R^2}) (r_+^2 - r_-^2)
\label{kappasq}
\ee
In our large $R$ limit we have  $a_i^2(\omega^2- \f{\lambda^2}{R^2}) \to 0$,  so  the angular equation reduces to the Laplacian on $S^3$. Thus
\be
\Lambda = l(l+2) + O(a_i^2(\omega^2- \f{\lambda^2}{R^2}))\label{Eqn:LambdaCorrections}
\ee

\subsection{The instability frequencies}

The radial equation cannot be solved exactly, but we can solve it in an `outer region' and an `inner region' and match solutions across the overlap of these regions.  We reproduce this computation of \cite{myers} in appendix \ref{one}. The instability frequencies correspond to the situation where there is no incoming wave, but we still have an outgoing wave carrying energy out to infinity. These instability frequencies are given by solutions to the transcendental equation
\be
-e^{-i \nu \pi} \f{\Gamma(1- \nu)}{\Gamma(1+ \nu)} \left(\f{\kappa}{2} \right)^{2\nu} = \f{\Gamma(\nu)}{\Gamma(-\nu)} \f{\Gamma(\h(1+ |\zeta| +\xi -\nu))\Gamma(\h(1+ |\zeta| -\xi -\nu)) }{   \Gamma(\h(1+ |\zeta| +\xi +\nu))\Gamma(\h(1+ |\zeta| -\xi +\nu))   }
\label{master}
\ee
We reproduce the solution to this equation, found in \cite{myers}, in appendix \ref{two}. 
In the large $R$ limit (\ref{limit}) the instability frequencies are real to leading order
\be
\omega \simeq \omega_R = \f{1}{R} \left(-l   - m_\psi m + m_\phi n - |-\lambda -m_\psi n + m_\phi m| -2(N+1) \right)
\label{omegare}
\ee
where $N\ge 0$ is an integer. The imaginary part of the frequency is found by iterating to a higher order; the result is
\be
\omega_I = \f{1}{R} \left(
  \f{2 \pi}{[l!]^2} \left[ (\omega^2-\f{\lambda^2}{R^2}) \f{Q_1Q_5}{4 R^2}\right]^{l+1}  \p^{l+1+N}C_{l+1} \p^{l+1+ N+ |\zeta|}C_{l+1} \right)
\label{omegaim}
\ee
Note that $\omega_I >0$, so we have an exponentially growing perturbation. Our task will be to reproduce (\ref{omegare}),(\ref{omegaim}) from the microscopic computation.

\section{The Microscopic Model: the D1-D5 CFT}
\setcounter{equation}{0}

In this section we discuss the CFT duals of the geometries of \cite{ross}. Recall that we are working with  IIB string theory compactified to $M_{4,1}\times S^1\times
T^4$.  The $S^1$ is parameterized by a coordinate $y$ with
\be
0\le y<2\pi R
\ee
The $T^4$ is described by 4 coordinates $z_1, z_2, z_3, z_4$. Let
the $M_{4,1}$ be spanned by $t, x_1, x_2, x_3, x_4$.  We have
$n_1$ D1 branes on $S^1$, and
$n_5$ D5 branes on
$S^1\times T^4$.  The bound state of these branes
is described by a 1+1 dimensional sigma model, with base space
$(y,t)$ and target space a deformation of the orbifold
$(T^4)^{n_1n_5}/S_{n_1n_5}$ (the symmetric product of $n_1n_5$ copies of $T^4$).  The
CFT has ${\cal N}=4$ supersymmetry, and a moduli space which
preserves this supersymmetry. It is conjectured that in this
moduli space we have an `orbifold point' where the target space is
just the orbifold
$(T^4)^{n_1n_5}/S_{n_1n_5}$ \cite{sw}.

The CFT with target space just one copy of $T^4$ is described by 4
real bosons
$X^1$,  $X^2$,  $X^3$,  $X^4$ (which arise from the 4 directions $z_1,
z_2,
z_3, z_4$), 4 real left moving fermions $\psi^1, \psi^2, \psi^3,
\psi^4$ and 4 real right moving fermions $\bar\psi^1, \bar\psi^2,
\bar\psi^3, \bar\psi^4$. The central charge is $c=6$.  The complete
theory with target space $(T^4)^{n_1n_5}/S_{n_1n_5}$ has $n_1n_5$ copies of this
$c=6$ CFT, with states that are symmetrized between the $n_1n_5$
copies. The orbifolding also generates `twist' sectors, which are
created by twist operators $\sigma_k$. A detailed construction
of the twist operators is given in  \cite{lm1,lm2}, but we summarize
here the properties that will be relevant to us.

The twist operator of order $k$ links together $k$ copies of the $c=6$
CFT so that the $X^i, \psi^i,\bar\psi^i$ act as free fields living on a circle
of length $k(2\pi R)$. Thus we end up with a $c=6$ CFT on a circle of length $k(2\pi R)$. 
We term each separate $c=6$ CFT a {\it component string}. Thus if we are in the completely untwisted sector, then we have $n_1n_5$ component strings, each giving a $c=6$ CFT
living on a circle of length $2\pi R$. If we twist $k$ of these component strings together by a twist operator, then they turn into one component string of length $k(2\pi R)$. In a generic CFT state there will be component strings of many different twist orders $k_i$ with $\sum_i k_i=n_1n_5$.

\subsubsection{The rotational symmetry $su(2)_L\times su(2)_R$}

The rotational symmetry of the noncompact directions $x_1\dots
x_4$ gives a symmetry
$so(4)\approx su(2)_L\times su(2)_R$, which is the R symmetry group of
the CFT. Let us start with the  left moving (L) sector. The 4 real fermions can be grouped into two complex fermions $\psi^+, \psi^-$ which form a doublet of  $su(2)_L$; the $U(1)$ part of this $su(2)_L$ is a $U(1)$ charge $j\equiv j^3_L$. Thus $\psi\equiv \psi^+$ and $\tilde\psi\equiv (\psi^-)^* $ will have $j={1\over 2}$, while $\psi^*, (\tilde\psi)^*$ will have $j ^*=-{1\over 2}$.
 Similarly, the fermions in the right moving sector  (R) can be grouped into a complex doublet which is a spin ${1\over 2}$ representation of $su(2)_R$. The $U(1)$  part of this group is $\bar j\equiv \bar j^3_R$. The fermions   $\bar\psi, \bar{\tilde\psi}$ will have $\bar j={1\over 2}$, while   $(\bar\psi)^*, (\bar{\tilde\psi})^*$ will have $\bar j=-{1\over 2}$.

\subsubsection{Chiral primaries and their descendants} 

The fermions of the CFT can be anti-periodic or periodic around the spatial circle $S^1$. If they  are antiperiodic,  we are in the Neveu-Schwarz  sector, and if they are periodic, we are in the Ramond  sector.

Let the CFT be in the  Neveu-Schwarz sector, and consider the left movers.  A special class of states are `chiral primaries' which have $h=j$. Similarly, we have `anti-chiral primaries', which have $h=-j$. Consider states which are described in both the left and right moving sectors by such primaries or their supersymmetry descendants. In the dual gravity description, these states correspond to the massless supergravity quanta.

In the CFT,   the chiral primary operators  are constructed by taking a twist operator $\sigma_k$ and adding a suitable charge; this construction was given in detail in \cite{lm2}. The simplest such operators $\sigma_k^{--}$ have scaling dimensions $h, \bar h$ and charges $j, \bar j$ given by
\be
\sigma_k^{--}: ~~h=j={k-1\over 2}, ~~\bar h=\bar j={k-1\over 2}
\ee

\subsection{ Spectral flow}

        The Neveu-Schwarz sector
states can be mapped to Ramond sector states by `spectral flow'
\cite{spectral}, under which the conformal dimensions and
charges change as
\bea
h'&=&h+\alpha j + \alpha^2{c\over 24}\\
j'&=&j+\alpha{c\over 12}
\label{qone}
\eea
Setting $\alpha=1$ gives a flow from the Neveu-Schwarz sector to the Ramond
sector. We see that under this flow anti-chiral primaries
of the Neveu-Schwarz sector
(which have $h=-j$) map to Ramond ground states with
\be
h={c\over 24}
\ee

\subsubsection{`Base spin'}

Suppose we have just one copy of the $c=6$ CFT, and this CFT is in the Neveu-Schwarz vacuum state. Then under spectral flow with $\alpha=1$ we would get a Ramond ground state with
\be
j={1\over 2}, ~~~\bar j={1\over 2}, ~~~h={1\over 4}, ~~~\bar h={1\over 4}
\ee
Thus we see that even though we are in a Ramond ground state, this state nevertheless has a spin under $su(2)_L\times su(2)_R$. The Ramond ground states form a representation $({1\over 2}, {1\over 2})$ under $su(2)_L\times su(2)_R$, so they can have $(j, \bar j)=(\pm {1\over 2}, \pm{1\over 2})$. We will call this value of $(j, \bar j)$ the `base spin' of the Ramond ground state, since we have this spin even before we have added any excitations. 

If we apply a twist operator $\sigma_k$ we link together $k$ copies of the  $c=6$ CFT to get a $c=6$ CFT on a circle of length $k(2\pi R)$. This long component string will have Ramond ground states which again have base spins in the representation $({1\over 2}, {1\over 2})$.

\subsubsection{Spectral flowed states}
 
The field theory on the physical D1-D5 branes that we have is in the Ramond sector.
This follows from the fact that the branes are solitons of
the gravity theory, and the fermions on the branes are induced
from fermions on the bulk. The latter are periodic around the
$S^1$; choosing antiperiodic boundary conditions would give a
nonvanishing vacuum energy and disallow the flat space solution
that we have assumed at infinity.

If we set $\alpha=2$ in (\ref{qone}) then we return to the Neveu-Schwarz
sector, and setting $\alpha=3$ brings us again to the Ramond sector.
More generally, the choice
\be
\alpha=2n+1\,, ~~~ n\in\mathbb{Z}
\label{qtwo}
\ee
brings us to the Ramond sector. Thus if we start with a simple state -- the vacuum -- in the Neveu-Schwarz sector, then we can obtain special excited states in the Ramond sector by spectral flow through $(2n_L+1), (2n_R+1)$ units in the left and right sectors. This is how we will obtain the states of our interest.

\subsection{The states we consider}

We will be working with the geometries constructed in \cite{ross}. Let us describe the CFT duals of these geometries. A subclass of these geometries will be extremal; these were constructed in \cite{gms1,gms2}, and their corresponding CFT states were identified. Here we will extend that analysis  to the more general geometries of \cite{ross}; the reader can refer to \cite{gms1,gms2} for more details of the CFT construction.

For the geometries of interest to us  the CFT state will be in the completely `untwisted' sector; i.e., we will have $n_1n_5$ separate copies of the $c=6$ CFT. As described above, the CFT is in the Ramond sector, so each copy of the $c=6$ CFT will be in some state of the Ramond sector. For our states of interest each copy of the CFT will in fact be in the same state. Further, this state is easily described. Recall that if we start with the Neveu-Schwarz vacuum and spectral flow by $\alpha=2n+1$ units, then we end up in a specific state of the Ramond sector. We will perform such a spectral flow by $\alpha=2n_L+1$ units for the left movers (L) and by $\alpha=2n_R+1$ units for the right movers (R). For each copy of the $c=6$ CFT we would get a state with
\be
j={(2n_L+1)\over 2}, ~~~\bar j={(2n_R+1)\over 2}, ~~~h={1\over 4}(2n_L+1)^2, ~~~\bar h={1\over 4}(2n_R+1)^2
\ee
Adding over all the $n_1n_5$ copies of the $c=6$ CFT we get
\be
j={(2n_L+1)\over 2}n_1n_5, ~~~\bar j={(2n_R+1)\over 2}n_1n_5, ~~~h={1\over 4}(2n_L+1)^2n_1n_5, ~~~\bar h={1\over 4}(2n_R+1)^2n_1n_5
\label{jjhh}
\ee
The angular momenta in the $\psi, \phi$ directions are given by
\be
J^{(cft)}_\psi = - j - \bar{j}, \qquad J^{(cft)}_\phi =  j - \bar{j}
\label{jq}
\ee
so for our state we get
\be
J^{(cft)}_\psi = - n_1 n_5 (n_R+ n_L +1)
\label{jpsicft}
\ee
\be
J^{(cft)}_\phi =  n_1 n_5 (n_L-n_R)
\label{jphicft}
\ee
The momentum of the state is given by $h-\bar h$. Note that the energies in the CFT are measured in units of ${1\over R}$, where $R$ is the radius of the circle on which the CFT lives. Thus the physical momentum of the state as seen from infinity will be
\be
P ={1\over R}[ h - \bar{h}] = {1\over R} [n_1 n_5 (n_L - n_R) ( n_L + n_R+1)]
\label{pp}
\ee
The extremal D1-D5 system in the Ramond sector arises if we  spectral flow the Neveu-Schwarz sector by $\alpha=1$ on both the L and R sectors. This state has
\be
h_{ex}={1\over 4} n_1n_5, ~~~~\bar h_{ex}={1\over 4} n_1n_5
\ee
Thus the physical energy above extremality of our {\it nonextremal} solution will be
\be
M-M_{ex}={1\over R}[h+\bar h-(h_{ex}+\bar h_{ex})]={1\over R}[n_L(n_L+1)+n_R(n_R+1)]n_1n_5
\label{ee}
\ee

Thus  the angular momenta, momentum and `mass above extremality' of the CFT states are given by (\ref{jpsicft}),(\ref{jphicft}),(\ref{pp}),(\ref{ee}) respectively. The geometries of \cite{ross} were labeled by  two integers $m,n$. Writing
\be
m=n_L+n_R+1, ~~~n=n_L-n_R
\label{nlnr}
\ee
we find that the angular momenta, momentum and `mass above extremality'  of the CFT states agree with the corresponding quantities (\ref{jpsival}),(\ref{jphival}),(\ref{pvalue}),(\ref{massgeom}) for the geometries. Note that the gravity quantities were computed after taking the large $R$ limit; only in this limit do we have a direct relation between gravity and CFT quantities.

\subsection{The fermionic excitations of the state}

We can give a more explicit description of the above CFT states. Since all copies of the $c=6$ CFT were in the same state, let us restrict to just one copy. There are no bosonic excitations $X^i$ in the state; the only excitations are due to fermions.  
Since we are in the Ramond sector, the fermions  are periodic around the spatial circle of the CFT and thus  have modes like
$\psi^i_n$, with $n$ integral. 

First consider the left sector. To get the state with spectral flow $n_L$ units we place one fermion $\psi$ in the lowest state with energy ${1\over R}$, the next in the level ${2\over R}$, and so on till we occupy the level ${n_L\over R}$. These fermions have an energy
\be
E={1\over R}[1+2+\dots n_L]={1\over R} {n_L (n_L+1)\over 2}
\ee
and a spin
\be
j={n_L\over 2}
\ee
The levels for the fermions $\tilde\psi$ are filled up in the same way upto the level ${n_L\over R}$. The total energy of the left movers is thus
\be
E_L={1\over R}n_L(n_L+1)
\ee
Recall that before we added any fermions the Ramond ground state already had a `base spin' of $j={1\over 2}$. After adding in the contributions of both fermions $\psi, \tilde\psi$ the spin is
\be
j=2({n_L\over 2})+{1\over 2}={2n_L+1 \over 2} 
\ee
Doing this for all the $n_1n_5$ copies of the CFT, and also adding the corresponding contributions of the right movers,  we see that we get the energy (\ref{ee}) and the charges given in (\ref{jjhh}).

In short, our microstates of interest have $n_1n_5$ copies of the $c=6$ CFT, with no twist operators linking the different copies of the CFT. We have fermionic excitations filling up the fermi sea with no `holes'. There are two species of such fermions for each of the left and right movers, and we have $n_L$ fermions of each species for the left movers and $n_R$ fermions of each species for the right movers. The state carries no bosonic excitations $X^i$. When we compute the emission from the microstates we will use the above explicit description of the state in terms of its excitations. 

\subsection{Twist operators}\label{which}

When a supergravity quantum is absorbed by the D1-D5 system then the process is represented in the CFT by insertion of a chiral primary or the supersymmetry descendent of a chiral primary. Such operators are composed of  a `twist' operator $\sigma_k$ and certain operators for bosonic and fermionic excitations. 

\begin{figure}[htbp] 
   \centering
   \includegraphics[width=3in]{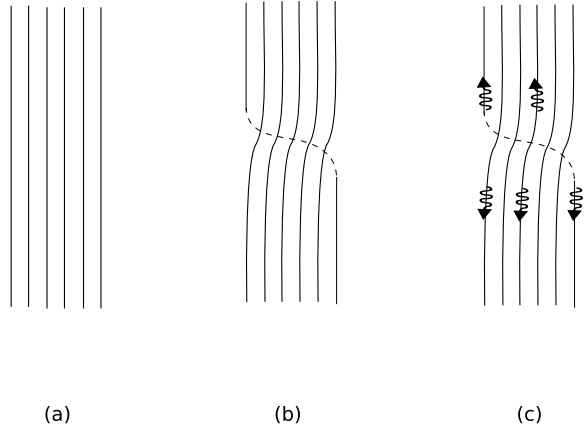} 
      \caption{(a) A subset of $k$ strands of the component strings near a point $y$ on the $S^1$ (b) A twist operator $\sigma_k$ inserted at $y$ changes the way these stands are linked; each strand joins the next one, in a cyclical fashion (c) The vertex operator for scalar emission also creates left and right moving excitations at the location $y$. }
   \label{Fig:twist}
\end{figure}

Let us describe the action of these operators in more detail. Suppose the supergravity quantum is incident at a point $(y_0,t_0)$ of the 1+1  dimensional CFT. There are $n_1n_5$ copies of the $c=6$ CFT in the vicinity of $(y_0,t_0)$; we are not at the moment concerned with how these copies link with each other as we go around the $y$ circle. The vicinity of $y_0$ is shown in fig.\ref{Fig:twist}(a). We have sketched only the $k$ copies of the $c=6$ CFT which will be involved in the twist operation. In fig.\ref{Fig:twist}(b) we show the result of the twist operation: the $k$ copies are linked with each other in a different way, with each copy turning into the next one as we cross the point $y_0$. In fig.\ref{Fig:twist}(c) we depict the fact that we also create some bosonic and fermionic excitations of the left and right movers at the location $(y_0, t_0)$. 

There are three things we need to know about the state created by the twist operator. 

\bigskip

(a) The  bosons and fermions can occupy {\it fractional modes}, i.e., the energy levels are ${n\over kR}$ ($n$ integer), rather than ${n\over R}$.

\bigskip 

(b) It should be noted however that when  we apply an operator to a state of the CFT, we do {\it not} necessarily end up creating a fractional level excitation. Thus suppose we have an operator $\partial X^i$ which creates a bosonic excitation. One might think that in the presence of the twist $\sigma_k$ the lowest allowed excitation will be in the energy level ${1\over kR}$. But consider again the $k$ strands shown in fig.\ref{Fig:twist}(a). If the operator $\partial X^i$  comes from a scalar incident on the D1-D5 state, then this scalar acts equally on all the $k$ strands. Suppose the $k$ strands were unlinked before the action of the twist operator. After the twist, they will form a single long loop of length $2\pi Rk$. The operator $\partial X^i$ will be inserted with equal amplitude at $k$ different points along this loop; these points are the $k$ images on the long loop of the point $y=y_0$. The action of any one of these insertions $\partial X^i$ would give rise to all harmonics
${n\over kR}$, but after we superpose the contributions of all the $k$ images of this insertion around the loop, we are left only with harmonics ${n\over R}$.  Similarly, consider the action  of the operator $J^+=\psi\tilde\psi$. By the same reasoning as above, we will excite only integer modes  $J^+_{-n}$. This will correspond to the fermions having fractional modes, but such that the total level is integral
\be
J^+_{-n}=\psi_{-m-{p\over k}}\tilde\psi_{m-n+{p\over k}}, ~~~n,m,p ~~{\rm integral}
\label{pptone}
\ee

\bigskip

(c) The last thing that we will need to know is the excitations of fig.\ref{Fig:twist}(c)  that correspond to the scalar that we will be considering. Our scalar arises from a graviton with both indices $i,j$ on the $T^4$. First consider absorption of the $l=0$ mode of this scalar into a D1-D5 state carrying no excitations. The energy of the scalar will  create an excitation \cite{dasmathur}
\be
{1\over \sqrt{2}}[\partial X^i\bar\partial X^j+\partial X^j\bar\partial X^i]\ee
In other words, we will create one left moving boson and one right moving boson; these carry the indices $i,j$. Now suppose we have $l>0$. Then in addition to the bosons we will create $l$ left moving fermions and $l$ right moving fermions. In the reverse process of {\it emitting} the scalar, excitations on the CFT can be annihilated to release the energy required for the emitted scalar. Thus we can annihilate one boson and $l$ fermions from the left movers, and one boson and $l$ fermions from the right movers, and create the desired scalar. But it will be important for us that this is not the only process that can create the scalar. An operator $\partial X$ in the CFT contains both creation and annihilation operators. Thus we can also emit the scalar for example by the following process: we annihilate   $l$ left and $l$ right fermions. and {\it create} a left and a right boson. 
More generally, we can create or annihilate any of the CFT excitations,  and the net energy released from the CFT becomes the energy of the emitted scalar.

\subsection{The vertex operator for scalar emission}\label{vertex}

With the above information, we can review the construction of the vertex operator in the CFT that corresponds to emitting a scalar. The properties of this operator were derived in \cite{mathurang} but we list the main steps here.

A bosonic quantum field has a normal mode expansion
\be
\hat \Phi=\f{1}{\sqrt{{\rm Vol}}} \sum_{\vec k} \left [~{1\over \sqrt{2\omega}} \left\{ \hat a_{\vec k}~e^{i(\vec k\cdot \vec x-\omega t)}~+~ \hat a_{\vec k}\dagger ~e^{-i(\vec k\cdot \vec x-\omega t)}\right \} ~ \right]
\label{ppten}
\ee
while a fermionic field has the expansion
\be
\hat\Psi=\f{1}{\sqrt{{\rm Vol}}} \sum_{\vec k} \left [~ \hat b_{\vec k}~e^{i(\vec k\cdot \vec x-\omega t)}~+~ \hat b^\dagger_{\vec k} ~e^{-i(\vec k\cdot \vec x-\omega t)}~ \right]
\label{ppel}
\ee
Here ${\rm Vol}$ is the volume of the space on which the fields live. In our problem we will assume that we have dimensionally reduced all fields on $T^4$. For fields in the 1+1 dimensional CFT we have ${\rm Vol}=2\pi R$. The emitted scalar however lives in the non-compact directions as well. Since we will be looking for radiation in a given partial wave, we expand the field operator for the scalar in polar coordinates. We regularize the wavemodes by choosing  a spherical box with a large radius $r_{max}$ and set wavemodes to  vanish at $r=r_{max}$. Then we find
\bea
\hat\Phi&=&\sum_{\omega, l, m_\psi,m_\phi} ~ {1\over \sqrt{2\pi R}}{1\over \sqrt{r_{max}}}\sqrt{\omega^3\over 2\pi}\Big ( {J_{l+1}(\omega r)\over (\omega r)}\Big ) \nn
&\times &\left [\hat a_{\omega, l, m_\psi,m_\phi}{1\over \sqrt{2\omega}}Y_{l, m_\psi, m_\phi}(\theta, \psi, \phi)e^{-i\omega t} +\hat a^\dagger_{\omega, l, m_\psi,m_\phi}{1\over \sqrt{2\omega}}Y^*_{l, m_\psi, m_\phi}(\theta, \psi, \phi)e^{i\omega t}\right ]\nn
\label{ppthone}
\eea
where 
\be
\omega={1\over r_{max}} {\pi\over 2} (2n+ l+ {1\over 2}), ~~~~(\omega > 0)
\label{ppthtwo}
\ee
We have normalized the wavemodes by noting that since $r_{max}$ is large, the leading contribution to the norm of the modes comes from the large $r$ domain where $J_{l+1}\approx \sqrt{2\over \pi\omega} r^{-{1/2}} \cos(\omega r -\pi/2(l+1)-\pi/4)$. (We have normalized $Y$ as ${1\over 2\pi^2}\int d\Omega_3 |Y|^2=1$.)

\bigskip

The interaction vertex has the following factors:

\bigskip

(i) From (\ref{ppthone}), the wavefunction of the scalar contributes a factor 
\be
{1\over \sqrt{2\pi R}}{1\over \sqrt{r_{max}}}\sqrt{\omega^3\over 2\pi}{1\over \sqrt{2\omega}}
\ee
where $\omega$ is the energy of the emitted scalar.

(ii) Let the left moving scalar $X$ in the CFT have an energy $\omega_1$. From (\ref{ppten}) we get a factor ${1\over \sqrt{2\pi R}}{1\over \sqrt{2\omega_1}}$. But  $X$ appears in the interaction through the term $\partial X$,  and the $\partial$ operator contributes a  factor $\omega_1$. With similar contributions from $\bar\partial X$ we get from these CFT scalars
\be
{1\over (2\pi R)}{\sqrt{\omega_1\bar\omega_1}\over 2}
\ee

(iii) There are $l$ left fermions and $l$ right fermions involved in the vertex. From (\ref{ppel}) we get the contribution 
\be
({1\over \sqrt{2\pi R}})^{2l}
\ee

(iv) The scalar wavefunction is $\sim J_l\sim r^l$ at the origin. This waveform has to interact with the CFT state placed at the origin $r=0$. If we wish to interact with the $l$th partial wave of the scalar, we need $l$ derivatives on   the scalar wavefunction. The left and right fermions carry spin ${1\over 2}$ under $su(2)_L$ and $su(2)_R$ respectively. Thus the interaction vertex has factors of the form
\be
\psi_L^\alpha \bar\psi_R^{\dot\alpha} \gamma^\mu_{\alpha\dot\alpha}\partial_\mu
\ee
and we see that with $l$ left and $l$ right fermions we have the correct number of derivatives $\partial_\mu$ on  the scalar wavefunction. But each application of this derivative brings in one factor of $\omega$, so we have a factor
\be
\omega^l
\ee

(v) The integral over the $S^1$ coordinate $y$ enforces momentum conservation along the $y$ direction, giving
\be
\int_0^{2\pi R} dy ~e^{i{n_1\over R} y}\dots e^{i{n_{l+1}\over R} y}e^{-i{\bar n_1\over R} y}\dots e^{-i{\bar n_{l+1}\over R}y}e^{-i{\lambda\over R} y} =(2\pi R) \delta_{\sum n_i-\sum\bar n_i- \lambda, 0}
\ee
where we have noted that there are $l+1$ left moving modes $n_i$, $l+1$ right moving modes $\bar n_i$, and the $y$ direction wavenumber of the emitted scalar is $\lambda$. We have $n_i>0$ is the left mover is annihilated, and $n_i<0$ if it is created; similarly for the other excitations. 

(vi) The creation and annihilation operators in the fields give a contribution depending on the excitations present in the appropriate modes. We write this contribution as
\be
{\mathcal D}(\omega_1){\mathcal D}(\omega_2)\dots {\mathcal D}(\omega_{l+1})
{\mathcal D}(\bar\omega_1){\mathcal D}(\bar\omega_2)\dots{\mathcal D}(\bar\omega_{l+1})\equiv \prod {\mathcal D}
\ee
Here $\omega_1$ is the energy of the left moving boson. The field operator for this boson can annihilate or create the boson. In the former case we will let $\omega_1>0$ and in the latter case $\omega_1<0$. For annihilation, if there are $n$ quanta already in the mode, we will get a factor ${\mathcal D}(\omega_1)=\sqrt{n}$. For creation, if there are $n$ quanta already in the mode, we have ${\mathcal D}(\omega_1)=\sqrt{n+1}$. The energies $\omega_2\dots \omega_{l+1}$ correspond to the $l$ left moving fermions. For fermion annihilation, if the mode is occupied ($n=1$) then ${\mathcal D}=0$, and if it is unoccupied ($n=0$) then ${\mathcal D}=1$. The right movers have the same physics; their energies are given by $\bar\omega_i$.

\bigskip

We will write the remaining factors in the vertex as a constant $V$. $V$ depends on the angular harmonics $l, m_\psi, m_\phi$, the charges $n_1, n_5$, the string tension $\alpha'$, the string coupling $g$ and   the volume of $T^4$. We will just write this constant as $V(l)$ to remind ourselves that $V$ depends on the angular quantum numbers;  the other parameters are fixed for the geometry. With all this, the amplitude for the transition per unit time is
\bea
{\cal R}&=&V(l)~\Big ({1\over \sqrt{(2\pi R)}} \f{1}{\sqrt{r_{max}}} \sqrt{\f{\omega^3}{2\pi}} {1\over \sqrt{2\omega}}\Big ) \Big ({1\over (2\pi R)}{\sqrt{\omega_1\bar\omega_1}\over 2}\Big ) \Big (({1\over \sqrt{2\pi R}})^{2l}\Big ) \nn
& &~~~\times ~~\Big (\omega^l
\Big )\Big((2\pi R) \delta_{\sum n_i-\sum \bar n_i- \lambda, 0}\Big)
\prod {\mathcal D}\nn
&=&V(l)~{1\over 4\pi} \sqrt{ \f{\pi}{r_{max}}} {1\over (2\pi R)^{l+{1\over 2}}}\omega^{l+1}\sqrt{\omega_1\bar\omega_1}\delta_{\sum n_i-\sum \bar n_i- \lambda, 0}
\prod {\mathcal D}
\label{rr}
\eea

We will use ${\cal R}$ in section (\ref{emission}) to compute the emission of scalars  from the CFT state, after determining $V(l)$ from the computation of semiclassical Hawking radiation.

\section{Obtaining $\omega_R$ from the CFT} \label{ob}
\setcounter{equation}{0}

In this section we describe the microscopic process of emission from the CFT. We will see that the emitted quanta have energies that agree with the  real part (\ref{omegare}) of the gravity modes.

In the gravity picture the  wavefunction of the emitted scalar was characterized by angular quantum numbers $l, m_\psi, m_\phi$. We need to relate these to the CFT description of these same quantities. Recall that in the CFT the angular $so(4)$ was written as $su(2)_L\times su(2)_R$. The scalar with angular harmonic $l$ is in the representation 
\be
({l\over 2}, {l\over 2})
\label{rep}
\ee
 of $su(2)_L\times su(2)_R$. The actual state in this representation is $(j, \bar j)$, with 
\be
-{l\over 2}\le j, \bar j\le {l\over 2}
\ee
Analogous to (\ref{jq}) we have the angular momenta in the $\psi, \phi$ directions
\be
m_\psi=-j-\bar j, ~~~m_\phi=j-\bar j
\label{repp}
\ee

\subsection{A simple example}

To illustrate the general emission process, let us begin with a very simple example. Recall that the state of the system was described by integers $n_L, n_R$ which gave` the number of units of spectral flow on the left and right sectors. We take the simplest non-extremal state, which has $n_L=n_R=1$. From (\ref{nlnr}) we see that the two integers describing the spectral flow of the gravity state are
\be
m=3, ~~ n=0
\label{mmnn}
\ee
In this state we have $n_1n_5$ copies of the $c=6$ CFT. In each copy of the CFT we have the following excitations.  In the left sector, we have one fermion $\psi$ and one fermion $\tilde\psi$, each in the energy level ${1\over R}$. The right movers similarly have fermions $\bar\psi, \bar{\tilde\psi}$ in the level ${1\over R}$. Thus the left and right energies for each component string are
\be
(E_L, E_R)=({2\over R}, {2\over R})
\ee
If this state emits a scalar, then the energy of the scalar has to be drawn from the energy carried by these fermionic excitations. 

Note that each copy of the CFT has a `base spin' $(j, \bar j)=({1\over 2}, {1\over 2})$ in $su(2)_L\times su(2)_R$, and the fermions contribute an additional spin $(1,1)$. Thus the spin of the state is 
\be
(j, \bar j)=({3\over 2}, {3\over 2})
\label{jjbars}
\ee
for each copy of the $c=6$ CFT.

\subsubsection{The emitted scalar: $l=2$}

Let us look for emission of a scalar carrying angular momentum $l=2$. Eq. (\ref{rep}) implies that the representation of $su(2)_L\times su(2)_R$ is $({{1}}, {{1}})$. With $m, n$ given by  (\ref{mmnn}), the energy $\omega$ of the emitted scalar is given by (cf. eq.(\ref{omegare}))
\be
\omega= {1\over R} \big [ -2   - 3m_\psi   - |-\lambda + 3 m_\phi | -2(N+1) \big ]={1\over R} \big [ -2   + 3(j+\bar j)  - |-\lambda + 3 (j-\bar j) | -2(N+1)\big ]
\label{omegareq}
\ee
To have emission, we need $\omega>0$. The only term on the RHS of 
(\ref{omegareq}) which can be positive is $3(j+\bar j)$.  Inspection of (\ref{omegareq}) shows that to get $\omega>0$ we {\it must} choose 
\be
j=1, ~~\bar j=1, ~~\lambda=0, ~~N=0
\label{set}
\ee
and with these parameters we get
\be
\omega={2\over R}, ~~~(j, \bar j)_{scalar}=({1}, {1})
\label{omom}
\ee
Our goal will now be to see how a scalar with exactly this energy and angular momentum is emitted by a transition between levels of the CFT.

\subsubsection{Energy of the quantum emitted from the CFT}

\begin{figure}[htbp] 
   \centering
   \includegraphics[width=3in]{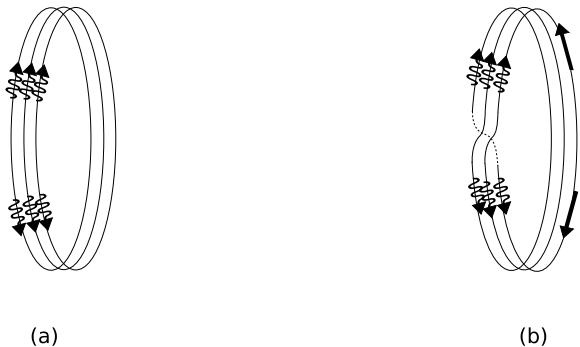} 
   \caption{(a) Three component strings, carrying their fermionic excitation   (b) The emission vertex changes these to one twisted component string.  The fermions now live on fractional energy levels on this longer component string, and a pair of bosons is created as well.  Angular momentum is lost because we have the  `base spin' of only one component string as opposed to three component strings; the scalar escapes with this angular momentum and the energy difference between the initial and final CFT states. 
}
   \label{fig:twoloops}
\end{figure}

Let us now see how a quantum with this frequency is emitted from the CFT state.  
 The operator that creates a quantum with $l=2$ includes the twist operator $\sigma_3$; i.e., it takes $3$ copies of the CFT, and twists them to one copy. In fig.\ref{fig:twoloops}(a) we depict three copies of the $c=6$ CFT state, with their fermionic excitations. Before the twist operator is applied (the `initial' configuration), the left and right energies carried by this set of three component strings is
 \be
 (E_L, E_R)_{initial}=3\times ({2\over R}, {2\over R})=({6\over R}, {6\over R})
 \label{eone}
 \ee
After the twist, we have one long  component string, on which the excitations can exist in units $({1\over 3R}, {1\over 3R})$ (fig.\ref{fig:twoloops}(b)). Let us take all the fermions in the initial untwisted state and transfer them to the twisted state. There will be three left fermions $\psi$, which will now have a total energy
\be
E={1\over R}[{1\over 3}+{2\over 3}+{3\over 3}]={2\over R}
\ee
With a similar energy for $\tilde\psi, \bar\psi, \bar{\tilde\psi}$, we find a total energy for the fermions in the `final' state
\be
(E_L, E_R)_{final}=({4\over R}, {4\over R})
\label{etwo}
\ee
One may think that since the energy (\ref{etwo}) is less than (\ref{eone}), we are done, and the difference will show up as the emitted energy of the scalar. But recall that the emission vertex also involves one left and one right boson $X$. These bosons carry the indices $i,j$ required of the 10-dimensional graviton $h_{ij}$ which is our emitted `scalar' in the theory dimensionally reduced on $T^4$. Since there are no bosons in the initial state, we must create them in the final state. Let us put these bosons in the lowest allowed level $({1\over R}, {1\over R})$. Then the difference between the energies of the initial  and the final states is
\be
(E_{L}, E_{R})_{initial}-(E_{L}, E_{R})_{final}=({6\over R}, {6\over R})-({4\over R}, {4\over R})-({1\over R}, {1\over R})=({1\over R}, {1\over R})
\ee
Thus the emitted scalar will carry an energy
\be
\omega_{emitted}=E_{L, emitted}+E_{R, emitted}={2\over R}
\ee
in agreement with (\ref{omom}).

\subsubsection{Spin of the  quantum emitted from the CFT} 

Let us now check the angular momentum of the emitted scalar. The spins carried by each component string (before twisting) is given by (\ref{jjbars}), so for the 3 component strings under consideration we have the  spin from the `initial' CFT state
\be
(j, \bar j)_{initial}=3\times ({3\over 2}, {3\over 2})=({9\over 2}, {9\over 2})
\label{jjabrt}
\ee
After the twisting, we have one component string. Let the  `base spin' of this twisted component string also be  $(j, \bar j)=({1\over 2}, {1\over 2})$. For the left movers, there are 3 fermions $\psi$ and 3 fermions $\tilde\psi$, contributing $j=3$ in all. Overall, the final state has
\be
(j, \bar j)_{final}=({7\over 2}, {7\over 2})
\ee
Thus the emitted scalar will have
\be
(j, \bar j)_{emitted}=(j, \bar j)_{final}-(j,\bar j)_{initial}=(1,1)
\ee
in agreement with (\ref{omom}).
 
 \subsubsection{Summary}
 
We thus see that the CFT state had just the right energy levels such that a transition between two of its levels produced a scalar with the correct energy and angular momentum to correspond to one of the emitted frequencies in the {\it gravity} calculation. The case of $l=2$ emission we described above is in a sense the simplest one possible, since a little inspection will show that we cannot have emission with $l=1$ in the geometry with  parameters (\ref{mmnn}). In the above discussion we made assumptions about how the excitations in the final state will behave; for example we assumed that all the fermions present in the initial state will show up in the final state in their lowest allowed levels, and that the base spin of the twisted component string was $(j, \bar j)=({1\over 2}, {1\over 2})$.  A little thought shows that these were indeed the only possibility in the present case. Suppose for example we {\it removed} one set of fermions $\psi, \bar \psi$ from the final state. This would seem to allow the emitted scalar to carry {\it more} energy. But then the spin quantum numbers of the emitted scalar would be $(j, \bar j)=({3\over2}, {3\over 2})$, which is not possible since for $l=2$ the $su(2)_L\times su(2)_R$ representation is
 $(1,1)$. 
 
We will now perform a more systematic analysis of different possible emissions. But a general underlying theme will be that in the simplest emission processes all the fermions on the initial component strings get transferred, with no change of spin, to the final twisted component string. Further, these fermions are placed in the lowest allowed state on this twisted string, so that they fill up a `fermi sea without holes'. More complicated processes can then be thought of as cases where we depart from this  simple case: there can be holes in the fermi sea, or changes in the spins of some fermions, or the creation of additional fermions.

\subsection{Emission of scalars with $m_\psi=-l, m_\phi=0$}\label{study}

The gravity calculation of instabilities showed that the frequencies of the emitted quanta are given by
\be
\omega ={1\over R} \big [ - l - m_\psi m + m_\phi n -2 -|\zeta| - 2N\big ]
\label{gravspec}
\ee
where $\zeta=-\lambda - m_\psi n + m_\phi m$. We wish to understand how such a spectrum emerges in the CFT description of emission. Let the emitted scalar have angular quantum number $l$. In this subsection we will consider the case where the emitted scalar has azimuthal quantum numbers $m_\psi=-l$; this implies $m_\phi=0.$ From (\ref{repp}) we see that this implies
\be
(j, \bar j)=({l\over 2}, {l\over 2})
\label{angq}
\ee
From the viewpoint of the 1+1 dimensional CFT, the energy $\omega$ and the momentum $\lambda$ of the scalar will arise from the energy and momentum released by the transition between CFT states. If the energies released from the left and right sectors are $\omega_L, \omega_R$ respectively then we will have 
\be
\omega = \omega_L+ \omega_R, \quad {\lambda\over R} = \omega_L - \omega_R
\label{olor}
\ee

\subsubsection{Spin quantum numbers}

The emitted scalar has angular quantum number $l$. Thus in the CFT the vertex operator will take $l+1$ component strings and twist then together into one long component string. Each component string has  a `base spin' $(j, \bar j)=({1\over 2}, {1\over 2})$. Further, it carries $n_L$ fermions $\psi$ and $n_L$ fermions $\tilde\psi$, each with $j={1\over 2}$. Similarly, each component string carries $n_R$ fermions $\bar\psi$ and $n_R$ fermions $\bar{\tilde\psi}$, each with $\bar j={1\over 2}$. Thus the   initial  $su(2)_L\times su(2)_R$ charge for all the $(l+1)$ component strings put together is
\be
(j, \bar j)_{initial} =\Big( (l+1)(n_L+\h), (l+1)(n_R+{1\over 2}) \Big)
\ee
The scalar carries $(j, \bar j)=({l\over 2}, {l\over 2})$, so the twisted component string in the final CFT state must carry a spin
\be
(j, \bar j)_{final}=\Big ((l+1)n_L+{1\over 2}, (l+1)n_R+{1\over 2}\Big )
\label{twistednum}
\ee

\subsubsection{Lowest energy state for the twisted component string}

Let us see what is the lowest energy state of the twisted component string which can carry the quantum numbers (\ref{twistednum}). We will first describe this state, and then argue that it is indeed the lowest energy state possible.

The `base spin' of the twisted component string is in the $su(2)_L\times su(2)_R$ representation $({1\over 2}, {1\over 2})$. Thus this spin can have values $(j, \bar j)=(\pm {1\over 2}, \pm {1\over 2})$. We will take $(j, \bar j)=({1\over 2}, {1\over 2})$. From (\ref{twistednum}) we see that the fermions on the twisted component string must then carry
\be
(j, \bar j)_{fermions, final}=\Big ((l+1)n_L, (l+1)n_R\Big )
\label{twistednumq}
\ee
We obtain this spin $j$ by taking $n_L$ fermions $\psi$ and $n_L$ fermions $\tilde\psi$, each of which has  $j={1\over 2}$, placed on the twisted component string in the lowest energy state allowed by the Pauli exclusion principle. Similarly, we obtain $\bar j$ by taking  $n_R$ fermions $\bar\psi$ and $n_R$ fermions $\bar{\tilde\psi}$, each with $\bar j={1\over 2}$,  in their lowest allowed energy state.

Now note that if we had taken for the base spin $j=-{1\over 2}$ instead
of $j={1\over 2}$, then we would need additional fermions to make up the required spin, and this would raise the energy above the minimum needed value.  Similarly, if we had taken some of the fermions to have $j=-{1\over 2}$ then we would again have needed extra fermions, leading to an increase in energy.

\subsubsection{Computing the energies}

First consider the energy of the component strings before twisting. There are $l+1$ component strings. For the left movers, on each component string we have $n_L$ fermions $\psi$ and $n_L$ fermions $\tilde\psi$, placed in the lowest modes allowed by the Pauli exclusion principle. The energy levels on these untwisted component strings are ${k\over R}$, with $k=1,2, \dots$. Thus we have for the left movers an energy
\be
E=2\times (l+1)\times {1\over R}[1+2+\dots n_L]=(l+1)n_L(n_L+1)
\ee
With a similar computation for the right movers,
\be
(E_L, E_R)_{initial}=\Big ({(l+1)n_L(n_L+1)\over R}, {(l+1)n_R(n_R+1)\over R}\Big )
\ee
Let us now compute the energy of this lowest energy state of the twisted component string. Consider the left movers. There are two species of fermions, $\psi$ and $\tilde\psi$, each separately satisfying the Pauli exclusion principle. The energy levels on the twisted component string are ${1\over R}{k\over l+1}$, with $k=1, 2, \dots$. We have $n_L(l+1)$ of each species of fermions. This gives for the left movers an energy
\be
E=2\times {1\over R(l+1)}[1+2+ \dots + n_L(l+1)]={1\over R}[n_L(n_L+1)(l+1)-n_L l]
\ee
We will let the bosonic excitation $X$ be in the lowest allowed mode ${1\over R}$. Thus the energies of the final state of the CFT are
\be
(E_L, E_R)_{final}=\Big ({n_L(n_L+1)(l+1)-n_L l+1\over R}, {n_R(n_R+1)(l+1)-n_Rl+1\over R}\Big )
\ee
Thus for the emitted scalar we find
\be
(\omega_L, \omega_R)=(E_L, E_R)_{initial}-(E_L, E_R)_{final}=\Big ( {n_Ll-1\over R}, {n_R l-1\over R} \Big )
\label{omomom}
\ee
From (\ref{olor}),
\be
\omega =  {1\over R} [(n_L+ n_R) l -2],  \qquad \lambda = (n_L - n_R) l
\label{omthree}
\ee
We can rewrite these in terms of the variables (\ref{nlnr}) appearing in the gravity construction
\be
\omega = {1\over R} [(m-1)l-2], \qquad \lambda= nl \label{omfour}
\ee
Recall that in the present case we have $m_\psi  = -l ,m_\phi=0$, so we can rewrite (\ref{omfour}) as
\be
\omega ={1\over R} [ -l - m_\psi m + m_\phi n - 2], \qquad \lambda=nl
\ee
Further, note that for the present case
\be
\zeta= -\lambda - m_\psi n = -nl + nl =0
\ee
So we can write $\omega$ in (\ref{omfour}) as 
\be
\omega ={1\over R} [ -l - m_\psi m + m_\phi n - 2 - |\zeta|]
\ee
We see that this agrees  with the spectrum (\ref{gravspec}) obtained from gravity when we set $N$ to its lowest value $N=0$.

\subsection{Obtaining general $m_\psi, m_\phi$}

So far we had considered the case where the angular quantum numbers of the scalar were $m_\psi=-l, m_\phi=0$, which by (\ref{repp}) corresponded to $(j, \bar j)=({l\over 2}, {l\over 2})$.  Let us now see how arbitrary values of $(j, \bar j)$ can arise. 

\subsubsection{The nature of excitations when $j<{l\over 2}$}

Suppose that $j={l\over 2}-1$. Then since the scalar carries one less unit of $j$, the final state of the twisted component string will have to carry one extra unit of $j$. This can be created by an application of $J^+\sim \psi\tilde\psi$, since each of $\psi, \tilde\psi$  carry $j={1\over 2}$. The exact details of this construction can only be understood if we make the full vertex operator for emission, following the lines of the constructions in \cite{lm1, lm2, lmpre}. We hope to return to this construction elsewhere, but for now will assume some basic properties of this vertex operator that appear reasonable in view of the constructions already seen in \cite{lm2, lmpre}. 

How much energy will the extra $\psi, \tilde\psi$ carry? The energy levels on the twisted component string are ${1\over R} {k\over l+1}$, with $k=1, 2, \dots$. In the case studied in the last section, we had fermions $\psi$ filled up to the level $k=n_L(l+1)$. One may thus think that we should place the $\psi, \tilde\psi$ in the levels with $k=n_L(l+1)+1$, getting an energy contribution from these two new fermions
\be
E_{\psi, \tilde\psi}=2\times {1\over R} {1\over l+1} [n_L(l+1)+1]={1\over R}[2n_L+{2\over l+1}] ~~~~~??
\ee
But as we noted in eq. (\ref{pptone}), the action of $J^+$ will create excitation energies in integer units of ${1\over R}$, though the individual excitations of the fermions $\psi, \tilde\psi$ can be fractional. Thus the lowest energy excitation can be obtained by taking $k=n_L(l+1)+1$ for $\psi$ and $k=n_L(l+1)+l$ for $\tilde\psi$. This will give an additional energy from these two fermions
\be
E_{\psi, \tilde\psi}={1\over R} {1\over l+1} [ n_L(l+1)+1+n_L(l+1)+l]={1\over R}(2n_L+1)
\label{eactual}
\ee
We could have obtained the same energy by taking $k=n_L(l+1)+2$ for $\psi$ and $k=n_L(l+1)+(l-1)$ for $\tilde\psi$. More generally, we can take $k=n_L(l+1)+s$, $s=1,2, \dots l$ for $\psi$ and $k=n_L(l+1)+(l+1-s)$ for $\tilde\psi$, so that we see that there are  $l$ such possibilities in all. Each time we decrease $j$ by one unit, we create some linear combination of these $l$ possible excitations. By the Pauli exclusion principle, we cannot use the same linear combination more than once, and so we can lower $j$ by upto $l$ units. But this corresponds to just the range covered by $j$: starting from $j={l\over 2}$, we can lower by one unit only $l$ times before we reach the lower bound $j=-{l\over 2}$.

We will assume this picture of excitation energies and compute the spectrum of the emitted scalar. We note however that this picture needs to be confirmed by an explicit construction of the vertex operator for scalar emission.

\subsubsection{Energy of the emitted scalar}

Let us compute the extra excitation energy that we have on the twisted component string of the final state, as compared to the corresponding energy that we had in the case $j={l\over 2}$. For $j<{l\over 2}$, we will need ${(l-2j)\over 2}$ units of the $\psi\tilde\psi$ excitations, each of which carry the energy (\ref{eactual}). This gives an extra energy
\be
{(l-2j)\over 2}\times {1\over R}(2n_L+1)
\ee
The energy of the emitted scalar will therefore be less by this amount as compared to the case $j={l\over 2}$. Doing a similar computation for the right movers, and using (\ref{omomom}), we see that we will have
\bea
(\omega_L, \omega_R)&=&\Big ( {n_Ll-1-(2n_L+1)({l\over 2}-j)\over R}, {n_R l-1-(2n_R+1)({l\over 2}-\bar j)\over R}\Big )\nn
&=&\Big ( {(2n_L)+1)j-{l\over 2}-1\over R}, {(2n_R+1)\bar j-{l\over 2}-1\over R}\Big )
\eea
This gives for the energy and momentum of the scalar
\be
\omega ={1\over R} [-l+  (2 n_L+1) j+(2 n_R+1) \bar j-2], \qquad  \lambda = (2 n_L+1)j-(2 n_R+1)\bar j
\ee
We can rewrite these in terms of the gravity variables
\bea
j &=&- \f{m_\psi - m_\phi}{2}, \qquad \bar j= -\f{m_\psi + m_\phi}{2} \nonumber \\
n_L &=& \f{m+n-1}{2} \qquad n_R= \f{m-n-1}{2}
\eea
to find
\be
\omega={1\over R} [-l - m m_\psi + n m_\phi -2], ~~~\lambda= m m_\phi - n m_\psi
\ee
We also have
\be
|\zeta| = - \lambda + m m_\phi - n m_\psi = 0
\ee
So we have
\be
\omega={1\over R} [-l - m m_\psi + n m_\phi -2 - |\zeta|]
\ee
which matches the gravity spectrum (\ref{gravspec}) for $N=0$.

\subsection{Obtaining nonzero $\zeta, N$}

In the computations of the above subsections we saw that in each case we had $\zeta=0, N=0$. The reason for this was that in each case we let the final state of the CFT be the lowest allowed energy state. We can change this state to be {\it not} the lowest energy state by
\begin{itemize}
\item
Exciting the fermions to higher energy levels, while maintaining the integrality constraints mentioned above. 
\item
Placing the bosons $X, \bar X$ in energy levels higher than the lowest one $({1\over R}, {1\over R})$ that we had chosen.
\item
Creating additional fermion pairs. For example we can create a left moving fermion $\psi$ with spin $j=-{1\over 2}$ and another fermion $\tilde\psi$ with $j={1\over 2}$. This keeps the overall spin unchanged but raises the energy of the state. We can also change the base spin of the twisted component string in the final state, which again needs the creation of additional fermions.
\end{itemize}

Making any such excitation will lower the energy available to the emitted scalar. Thus we can write
\be
\omega_L = {1\over R} [(2 n_L+1) j - \f{l}{2} -1 - \delta_L], ~~~\omega_R ={1\over R} [ (2 n_R+1) \bar j - \f{l}{2} -1 - \delta_R]
\ee
This gives for the energy and the momentum of the scalar
\be
\omega ={1\over R} [-l+  (2 n_L+1) j+(2 n_R+1) \bar j-2 - \delta_L - \delta_R], \qquad  \lambda = (2 n_L+1)j-(2 n_R+1)\bar j + \delta_R - \delta_L
\ee
where $\delta_L,\delta_R$ are both positive. Expressing these results in terms of the gravity variables gives
\be
\omega={1\over R} [-l - m m_\psi + n m_\phi -2 - \delta_L - \delta_R]
\ee
and
\be
\lambda = m m_\phi - n m_\psi+ \delta_R - \delta_L
\ee
We can further write
\be
\zeta=\delta_L - \delta_R
\ee
If $\zeta >0$ we can write the energy of the emitted scalar as
\be
\omega ={1\over R} [ -l - m m_\psi + n m_\phi -2 - \zeta - 2 \delta_R]
\ee
and if  $\zeta <0 $ we can write the energy as
\be
\omega ={1\over R} [ -l - m m_\psi + n m_\phi -2 +\zeta -2 \delta_L]
\ee
In either case the energy has the form
 \be
\omega ={1\over R} [ -l - m m_\psi + n m_\phi -2 -|\zeta| -2 N]
\label{ppttwo}
\ee
where $N\ge 0$. If we assume, following the arguments in section \ref{which}, that the energies of the left and right sectors are integral, then we have $N=0,1,2\dots$, and the expression (\ref{ppttwo})  agrees with the gravity expression (\ref{gravspec}).

\section{Relating $\omega_I$ to the radiation rate} \label{radiationrate}
\setcounter{equation}{0}

The classical instability observed in the gravity solution is described by a complex frequency $\omega=\omega_R+i\omega_I$. The real part $\omega_R$ gives the energy of the emitted scalars, and we saw in the above section that the spectrum of  $\omega_R$ was reproduced in the CFT computation. We will now see how $\omega_I$ is related to the {\it rate} of emission of these scalar quanta from the CFT state.

Radiation from the CFT state is of course not a new effect. Consider the near-extremal D1-D5 system. The entropy of CFT excitations agrees with the Bekenstein entropy of the corresponding extremal and near-extremal hole \cite{Strominger:1996sh,callanmalda}.  To study Hawking emission from the near-extremal hole we must look at a generic state of this near-extremal D1-D5 CFT. The bosonic and fermionic excitations  in this state are described by thermal distributions $\rho_B, \rho_F$. Emission from these excitations reproduces all the properties of the low energy Hawking radiation from the corresponding black hole: spin dependence, emission rates and greybody factors \cite{callanmalda,dasmathur,maldastrom}. 

But this radiation is a quantum phenomenon, obtained on the gravity side by  a semiclassical computation of quantum fields on the curved black hole background. It may therefore seem surprising that for the geometries of \cite{ross} that we are studying the emission manifests itself as a {\it classical} instability. Why does  the  CFT computation produce a stronger effect in our present case?

The reason for the classical nature of the instability lies in the fact that we are considering emission from a particular state, which is not a very `generic' state. In this CFT state we have $n_1n_5$ component strings which are all identical: they are all untwisted (`singly wound') and they all carry the same excitations. By contrast in a generic state of the D1-D5 system there will be component strings of different twist orders, carrying different excitations. Thus  while the underlying emission process will be the same in our particular state and in the generic state, the details of the emission can be very different. Let us now see how these differences will arise.

Note that each component string carries an equal number of left and right fermions: thus it is `bosonic'. (The base spin $({1\over 2}, {1\over 2})$ is also bosonic overall, since it is ${1\over 2}$ on each side.) The final CFT state (the twisted component string)  is also bosonic; this follows since the emitted quantum was bosonic. Whenever we have identical bosons in a state, we encounter the phenomenon of `bose enhancement'. Let us summarize how this will work in our present case.

\bigskip

(a) The  component strings present in the initial state are all identical bosons. In an amplitude involving such a component string there will thus be a factor  $\sim \sqrt{N}$ where $N$ is the number of these component strings. But  $N$ is large to start with and we will evolve the perturbation only for short times so that it will not change significantly; this corresponds to the fact that  in the dual gravity calculation we evolve the perturbation only for short times so that backreaction on the geometry can be ignored. Thus even though $N$ will decrease in principle during the evolution, we will not see a change in the factor $\sqrt{N}$ at the order where we work.

(b) The emission of scalars will lead to the creation  of  `twisted' component strings. After $n$ scalar quanta have been emitted, there will be $n$ such twisted component strings created. The creation of the next twisted component string will contribute a factor $\sqrt{n+1}$ to the amplitude, and thus a factor $n+1$ to the probability of emission. This will be the important effect: the rate of radiation will keep increasing as we emit more and more quanta, because more and more twisted component strings get created `in the same state' and bose enhancement follows.

(c) The emitted scalars are also bosons, and in a system like a laser these could have a bose enhancement factor. But to get such a factor we would have to confine these scalars by some potential which would reflect them back to the emitting system. We do not have such a potential here, and the emitted scalars just escape to infinity. Thus we do not get a bose enhancement factor from the emitted scalars.

\bigskip

We will discuss the bose enhancement factors in more detail below, but the nature of the final relation can be seen easily. Suppose we have emitted $n$ quanta, and the probability of emission of the $(n+1)^{\text{st}}$ quantum is $(n+1)$ times the probability of emitting the first one. Then
\be
{dn\over dt}=\mu (n+1)
\label{ppsev}
\ee
for some constant $\mu$. If we find the probability of emission per unit time for the first quantum, then we will have
\be
\Gamma={dn\over dt}|_{n=0}=\mu
\label{pptfive}
\ee
On the other hand, in the classical limit where a large number $n$ of quanta have been emitted we will have
\be
{dn\over dt}\approx \mu n, ~~~n(t)\sim e^{\mu t}
\ee
The number of quanta $n$ in the classical  scalar perturbation  $\phi$ is proportional to the square of the field $\phi$
describing the perturbation
\be
n\sim |\phi|^2
\ee
Thus
\be
|\phi(t)|\sim e^{{\mu\over 2} t}
\label{pptfour}
\ee
But we have described the growth of the perturbation by the complex frequency $\omega=\omega_R+i\omega_I$, so the growth of $\phi$ is given by
\be
|\phi|\sim e^{\omega_I t}
\label{pptthree}
\ee
Comparing (\ref{pptthree}) and (\ref{pptfour})  we see that if we compute the rate of emission $\Gamma$ of the first quantum from the CFT state (\ref{pptfive}) then we will have
\be
\Gamma=2\omega_I
 \label{imag}
\ee
We will see in the next section that this relation is exactly satisfied.

\subsection{Emission rates and Bose-Einstein condensates (BECs)}

In the process of radiation from our CFT state we have seen that we start with a set of untwisted component strings, carrying some excitations, and these turn into a twisted component string with a lower net energy. The difference in energy shows up as the energy of the emitted scalar. The essential part of the physics can be modeled by the following simple system. Suppose we have a large number $N$ of atoms, which are all in the {\it same} excited state. These atoms can decay to a lower energy state with the emission of a photon. Thus after $n$ photons have been emitted, we have $n$ atoms in the lower energy state, but note that these are also identical particles, and we will have to use bose statistics to describe them. In short, we have a Bose-Einstein condensate (BEC) in the excited state; this condensate can decay to a BEC in the unexcited state, with the emission of photons. We assume that the photons themselves escape the system since they are radiated into a noncompact space. In this situation, how does the emission rate behave as a function of time?

\subsubsection{The interaction Hamiltonian}

First consider one atom, and just one fourier mode of the photon. The interaction Hamiltonian leading to the emission can be written as
\be
H_{int}=\alpha \hat O \hat a^\dagger + \alpha^* \hat O^\dagger \hat a
\ee
Here $\alpha$ is a constant giving the strength of the interaction, $\hat O$ is an operator that changes the excited state of the atom to the unexcited state, and $\hat a^\dagger$ creates the photon. If there are $N$ atoms, then we will have
\be
H_{int}= \alpha (\sum_{i=1}^N \hat O_i) \hat a^\dagger + \alpha^* (\sum_{i=1}^N \hat O^\dagger_i)\hat a
\label{ppsix}
\ee
A general state of this system is described by giving the number of atoms $n$ that are in the `de-excited' state, and the number $k$ of photons in the chosen fourier mode. (Of course if we started with all atoms excited and no photons, then $n=k$ during the evolution, but we can leave $n, k$ arbitrary for now.) Let this state be denoted
\be
|n, k\rangle
\ee
We will have
\be
H_{int}|n, k\rangle=A|n+1, k+1\rangle+B|n-1, k-1\rangle
\ee
where $A,B$ are constants; our goal is to find $A, B$.

The action of $\hat a^\dagger$ is obvious
\be
\hat a^\dagger |n, k\rangle=\sqrt{k+1}|n, k+1\rangle
\ee
The action of $\sum_i \hat O_i$ needs a little more thought, though the final result has a simple interpretation. Let us write the $N$ atoms with distinct labels, but making sure that the wavefunction is symmetric under any permutation of the atoms. Then we have
\be
|n,k\rangle={1\over \sqrt{{}^N C_n}}\left[ |\underbrace{\downarrow\dots\downarrow}_n\underbrace{\uparrow\dots\uparrow}_{N-n} ~ \rangle +{ permutations} \right]
\label{ppone}
\ee
where $\downarrow$ represents a de-excited atom and $\uparrow$ an excited atom.
On the RHS we have written explicitly the term where the first $n$ atoms are in the de-excited state, but then noted that we have to include all permutations which will make different sets of atoms be in the de-excited state. There are ${}^N C_n$ such terms in all, and the prefactor makes the state normalized
\be
\langle n, k|n, k\rangle =1
\ee

We wish to find the action of $\sum_i \hat O_i$ on this state. First consider the action of $\sum_i\hat O_i$ on $|\underbrace{\downarrow\dots\downarrow}_n\underbrace{\uparrow\dots\uparrow}_{N-n}~\rangle$. For $i=1,\dots n$ we get zero, since the corresponding atoms are already de-excited.  For each value of $i=n+1,\dots N$ we get a term with $n+1$ atoms de-excited. Thus we create $N-n$ such terms. In (\ref{ppone}) we had ${}^N C_n$ terms in the square bracket on the RHS. The action of $\sum_i \hat O_i$ on this set of terms will generate
\be
{}^N C_n \times (N-n)
\ee 
terms, in each of which we have $n+1$ de-excited atoms. 

But many of these terms will have to be identical to each other, since there are only
\be
{}^N C_{n+1}
\ee
different ways to select which $n+1$ atoms will be de-excited. Thus the state with any given set of atoms de-excited will appear in our sum
\be
{{}^N C_n \times (N-n)
\over {}^N C_{n+1} }
\ee 
times. Thus we find
\be
(\sum_i \hat O_i)|n, k\rangle = \Big({1\over \sqrt{{}^N C_n}}\Big )\Big({{}^N C_n \times (N-n)
\over {}^N C_{n+1} }\Big )\left[ |\underbrace{\downarrow\dots\downarrow}_{n+1}\underbrace{\uparrow\dots\uparrow}_{N-n-1}~\rangle+{ permutations} \right]
\ee
The number of terms in the square bracket on the RHS is $ {}^N C_{n+1} $, and each of these terms is orthogonal to the others since they all correspond to different sets of atoms being de-excited. We can therefore write the terms in the square bracket in terms of a normalized state
\bea
(\sum_i \hat O_i)|n, k\rangle &=& \Big({1\over \sqrt{{}^N C_n}}\Big )\Big({{}^N C_n \times (N-n)
\over {}^N C_{n+1} }\Big )\Big( \sqrt{{}^N C_{n+1}}\Big )\nn
&&~~~~~~~~~~~~~~~~~\times ~~{1\over \sqrt{{}^N C_{n+1}}}\left[ |\underbrace{\downarrow\dots\downarrow}_{n+1}\underbrace{\uparrow\dots\uparrow}_{N-n-1} ~\rangle+{ permutations} \right]\nn
\eea
But
\be
 \Big({1\over \sqrt{{}^N C_n}}\Big )\Big({{}^N C_n \times (N-n)
\over {}^N C_{n+1} }\Big )\Big( \sqrt{{}^N C_{n+1}}\Big )=\sqrt{N-n}\sqrt{n+1}
\ee
and
\be
{1\over \sqrt{{}^N C_{n+1}}}\left[ |\underbrace{\downarrow\dots\downarrow}_{n+1}\underbrace{\uparrow\dots\uparrow}_{N-n-1} ~\rangle+{ permutations} \right]=|n+1, k\rangle
\ee
Thus
\be
(\sum_i\hat O_i)|n, k\rangle=\sqrt{N-n}\sqrt{n+1}|n+1, k\rangle
\label{ppfour}
\ee
Similarly, we find
\be
(\sum_i\hat O^\dagger_i)|n, k\rangle=\sqrt{N-n+1}\sqrt{n}|n-1, k\rangle
\label{ppfive}
\ee
and overall we get
\be
H_{int}|n, k\rangle=\alpha \sqrt{N-n}\sqrt{n+1}\sqrt{k+1}|n+1, k+1\rangle+\alpha^*\sqrt{N-n+1}\sqrt{n}\sqrt{k-1}|n-1, k-1\rangle
\label{ppeig}
\ee
This relation has a simple interpretation. The state $|n, k\rangle$ had $N-n$ atoms in the excited state and $n$ atoms in the de-excited state. We can regard the action of $\sum_i \hat O_i$ as annihilating an excited atom and creating a de-excited atom.  Since the atoms in each state behave as identical bosons, we get a factor $\sqrt{N-n}$ from annihilating the excited atom, and a factor $\sqrt{n+1}$ from creating a new de-excited atom. Introducing creation and annihilation operators $\hat A_\uparrow^\dagger, \hat A_\uparrow^{\phantom \dagger}$ for the excited atom, and creation and annihilation operators  
$\hat A_\downarrow^\dagger, \hat A_\downarrow^{\phantom \dagger}$ for the de-excited atom
\be
[\hat A_\uparrow^{\phantom \dagger}, \hat A_\uparrow^\dagger]=1, ~~~[\hat A_\downarrow^{\phantom \dagger}, \hat A_\downarrow^\dagger]=1
\ee
we can write
\be
\sum_i\hat O_i=\hat A^\dagger_\downarrow \hat A_\uparrow^{\phantom \dagger}, ~~~\sum_i\hat O^\dagger_i=\hat A^\dagger_\uparrow \hat A_\downarrow^{\phantom \dagger}
\ee
Thus we see that if we introduce field operators for the excited and de-excited atoms then the factors in (\ref{ppfour}),(\ref{ppfive}) become obvious; they are just the usual bose factors for these states of the atoms.

\subsubsection{Applying the model to our system}

Let us now see how the above model of atoms/radiation applies to our problem:

\bigskip

(i) In place of the excited atoms we will have the untwisted component strings present in the
initial state of our CFT. These component strings carry left and right moving fermions, and a part of the energy of these fermions will be released to radiation. The process of radiation will take $l+1$ of these component strings and twist them together. In our toy model each atom could radiate by itself, so for $l>0$ the combinatorics of decay will be slightly different between our system and the atom model. But this difference is not material for our discussion, because we will see that the  effects of interest do not depend on the number of quanta in the excited state. In the atom model, we started with $N$ atoms in the excited state, and after some time $t$ we have $N-n(t)$ atoms in the excited state. The amplitude for emission of the next quantum has a factor $\sqrt{N-n(t)}$. But we start with $N\gg 1$, and we evolve for times only such that $n(t)<<N$; this restriction corresponds to the fact that in our gravity system we ignore the backreaction of the emitted scalar on the geometry. Thus we will have
\be
\sqrt{N-n(t)}\sim \sqrt{N}
\ee
and so the radiation rate is not affected to leading order by the change in the number of atoms in the excited state. For the same reason, in our actual problem the combinatorics of the component strings in the initial state will give a contribution which we can take as a constant in time (we will compute this constant in the next section).

(ii) In the toy problem we see that after we have $n$ atoms in the de-excited state, the amplitude for de-exciting the next atom has a factor $\sqrt{n+1}$; thus the probability for de-excitation has a factor $n+1$. In the actual CFT problem, each time we emit a scalar we create a twisted component string. These twisted component strings are the analog of the de-excited atoms. We see that we get the relation (\ref{ppsev}) governing the evolution of $n(t)$. Thus the exponential growth of the perturbation results from the fact that if there are already $n$ twisted component strings in the CFT state then `induced emission' makes it easier to create the next twisted component string.

(iii) The photons in the toy model are the analogs of scalars in the actual problem. These quanta are radiated into noncompact space. If we regularize the volume of this noncompact space by taking it to be  a large ball with radius $r_{max}$, then we will have a factor ${1\over \sqrt{r_{max}}}$ in the coupling $\alpha$ in (\ref{ppsix}). Thus the amplitude for emission into any given fourier mode of this massless quantum will be small, and the finite rate of emission will result from the fact that the number of levels grows as $\sim r_{max}$, forming a band when $r_{max}\rightarrow\infty$. We will thus use the Fermi golden rule in the next section to compute radiation into this band, but note that since the occupation number of each individual mode is small there is no Bose enhancement from the factor $\sqrt{k+1}$ in 
 (\ref{ppeig}).
 
Note that in a laser one places mirrors at the end of the emitting medium to reflect the emitted photons back into the system; this populates the photon modes to levels $k\gg 1$ and gives a Bose enhancement for emission of further quanta. In our problem we could imagine creating a potential that would reflect the emitted scalars back towards the origin, as in the example of the `black hole bomb' \cite{cardoso}. It would be interesting to see the analogous computation in the present problem.

\bigskip

\subsection{The waveform of the emitted field}

In the gravity computation we compute the scalar field perturbation as a function of space and time $\phi=\phi(\vec r, t)$. We find that $\phi$ grows exponentially with time, and falls exponentially  with $|\vec r|$. In the CFT computation we  think of this scalar field as arising from scalar quanta emitted from the CFT, and these quanta can escape to infinity. We will see that the rate of emission of these quanta increases exponentially at the same rate as the scalar waveform in the gravity calculation, and that will establish the equivalence we seek between the gravity and CFT computations.  But in this subsection we give a qualitative picture of why the exponentially growing emission from the CFT leads to a waveform for the scalar with exponential behaviors in both $|\vec r|$ and $t$. 

The essential physics is that the quanta emitted from the CFT state at $\vec r=0$ `pile up' at finite $|\vec r|$; i.e., they are produced at a rate faster than the rate at which they can flow off to infinity. This leads to a growth in the number density of quanta everywhere. We illustrate this with a toy model.

Consider a lattice of points, labeled by $i=0, 1, 2, \dots$. Scalar quanta are produced at the origin $i=0$, and they can hop, one site at a time, to $i=1,2,\dots$, thus escaping towards $i\rightarrow \infty$. At time $t$ let there be $n_i(t)$ quanta at site $i$. At $i=0$ we produce quanta at a rate which is proportional to the number of quanta already emitted; this accords with what we found in the above subsections about the nature of our system. Suppose $\tilde n(t)$ quanta have been emitted by time $t$. Then 
\be
{d\tilde n(t)\over dt}=\mu ~\tilde n(t)
\label{pptsev}
\ee
where $\mu$ is a constant. This has the solution
\be
\tilde n(t)=Ce^{\mu t}
\ee
The rate at which quanta hop from site $i=0$ to $i=1$ is proportional to the number of quanta $n_0(t)$ at $i=0$.  Thus we have
\be
{dn_0(t)\over dt}=\mu ~\tilde n(t)-\alpha_0 n_0(t)
\label{ppnine}
\ee
where the first term comes from the new quanta being produced,  and $\alpha_0$ is a constant governing the hopping from $i=0$ to $i=1$. This relation gives
\be
n_0(t)={C\mu\over( \mu+\alpha_0)}e^{\mu t}
\ee
In the same way, for the number of quanta at the site $i=1$ we will have
 \be
 {dn_1(t)\over dt}=\alpha_0 n_0(t)-\alpha_1 n_1(t)
 \ee
 where $\alpha_1$ is a constant the governs hopping from $i=1$ to $i=2$. We find
 \be
 n_1(t)={\alpha_0\over (\mu+\alpha_1)}~{C\mu\over( \mu+\alpha_0)}e^{\mu t}
 \ee
 and more generally
 \be
 n_i(t)=\prod_{j=0}^i \Big({\alpha_{j-1}\over (\mu+ \alpha_j)}\Big) Ce^{\mu t}
 \ee
 where we have set $\mu=\alpha_{-1}$. If we assume that for some $i>i_0$ all the $\alpha_i$ have the same value $\bar\alpha$, then we will have
 \be
{n_{i+1}\over n_i}={\bar\alpha\over \mu+\bar\alpha}<1 ~~~~(i>i_0)
\ee
We see that $n_i$ falls off exponentially as $i\rightarrow \infty$. We have already seen that all the $n_i(t)$ grow exponentially in time with the behavior $\sim e^{\mu t}$. These are just the properties of the scalar perturbation in the gravity solution. Thus we see that if quanta are produced with a rate given by 
 (\ref{pptsev}) and the allowed to disperse then they will produce a profile of the scalar field of the type seen in the gravity calculation.

\subsection{Summary}

In this section we have tried to give a qualitative picture of how emission from the CFT state placed at $r=0$ reproduces a scalar waveform of the kind seen in the classical instability. The key relation that we need in what follows is (\ref{imag}). This relation tells us that the exponential rate of growth of the perturbation, given by $\omega_I$, is related to the rate $\Gamma$ of spontaneous emission of the first quantum from the system (i.e., before bose enhancement sets in). In the following section we will compute $\Gamma$, by taking the CFT state where all component strings are in the excited state, and looking at the process where $l+1$ component strings get converted to the 
twisted state.

\section{The rate of radiation}\label{emission}
\setcounter{equation}{0}

The classical instability of \cite{myers} has an imaginary $\omega_I$ part to its frequency, and we have seen in (\ref{imag}) that the rate of radiation from the CFT state should be given by $\Gamma=2\omega_I$. In this section we will compute $\Gamma$ and see that this relation is indeed true. We will focus of the emission  process that we studied in section \ref{study}:
\be
m_\psi=-l, ~~~m_\phi=0, ~~~\lambda=0, ~~~\zeta=0, ~~~N=0
\ee
For these parameters, $\omega_I$   given by (\ref{omegaim}) equals
\be
\omega_I = \f{1}{R} \left(
  \f{2 \pi}{[l!]^2} \left[ \omega^2 \f{Q_1Q_5}{4 R^2}\right]^{l+1} \right)
\label{omegaimq}
\ee

We will compute $\Gamma$ in three steps:

\bigskip

(a) We will compute the semi-classical Hawking radiation $\Gamma_l$ from a D1-D5 black hole.

(b) We look at the CFT model for this emission, and observe that the $\omega$ dependence of $\Gamma_l$ is reproduced by the CFT computation, upto an overall normalization constant that we do not fix from the CFT. The computation assumes that the CFT quanta are distributed `thermally', with bose and fermi distributions $\rho_B, \rho_F$. We use the semiclassical hawking radiation rate to fix the overall constant $V(l)$ needed in the CFT computation.

(c) In the CFT computation we replace the distributions $\rho_B, \rho_F$ with the actual occupation numbers for CFT levels that are correct for our chosen microstate. With these new occupation numbers we recompute $\Gamma_l$ from the CFT and observe that this new $\Gamma_l$ equals $2\omega_I$.

\subsection{The semiclassical Hawking radiation}

To compute the semiclassical Hawking radiation from a hole, one computes the absorption cross section $\sigma$ and then uses detailed balance to get the emission.

The classical absorption cross section is, for $l$ odd \cite{mathurang}
\bea
\sigma_l&=&(l+1)^2 {\pi^3}{(Q_1Q_5)^{l+1}\over 2^{4l}[l!(l+1)!]^2}
\omega^{2l-1}\nn
&&~~~~~~\times [\omega^2 +(2\pi T_L)^2 1^2][\omega^2 +(2\pi T_L)^2 3^2]\dots [\omega^2 +(2\pi T_L)^2 l^2]\nn
&&~~~~~~\times [\omega^2 +(2\pi T_R)^2 1^2][\omega^2 +(2\pi T_R)^2 3^2]\dots [\omega^2 +(2\pi T_R)^2 l^2]\nn
&&~~~~~~\times {e^{\omega\over T_H}-1\over (e^{\omega\over 2 T_L}+1)(e^{\omega\over 2 T_L}+1)}
\eea
and for $l$ even
\bea
\sigma_l&=&(l+1)^2 {\pi^3}{(Q_1Q_5)^{l+1}\over 2^{4l}[l!(l+1)!]^2}
\omega^{2l+1}\nn
&&~~~~~~\times [\omega^2 +(2\pi T_L)^2 2^2] [\omega^2 +(2\pi T_L)^2 4^2]\dots [\omega^2 +(2\pi T_L)^2 l^2]\nn
&&~~~~~~\times[\omega^2 +(2\pi T_R)^2 2^2] [\omega^2 +(2\pi T_R)^2 4^2]\dots [\omega^2 +(2\pi T_R)^2 l^2]\nn
&&~~~~~~\times {e^{\omega\over T_H}-1\over (e^{\omega\over 2 T_L}-1)(e^{\omega\over 2 T_L}-1)}
\eea
Let $\Gamma_T$ be the total number of quanta emitted per unit time. Then
\be
d \Gamma_T= \sigma {d^4k\over (2\pi)^4}{1\over e^{\omega\over T_H}-1}=\sigma {(2\pi^2)\omega^3 d\omega\over (2\pi)^4}{1\over e^{\omega\over T_H}-1}
\ee
where $2\pi^2$ is the area of the unit 3-sphere. The scalar with angular momentum $l$ has $(l+1)^2$ polarizations, each of which is emitted with equal probability since the hole has no angular momentum. So number of quanta with given polarization emitted per unit time is given by $\Gamma=\Gamma_T/(l+1)^2$.  We find, for $l$ odd
\bea
d \Gamma_l&=&{\pi\over 8}{(Q_1Q_5)^{l+1}\over 2^{4l} [l!(l+1)!]^2}
\omega^{2l+2}d\omega\nn
&&~~~~~~\times [\omega^2 +(2\pi T_L)^2 1^2][\omega^2 +(2\pi T_L)^2 3^2]\dots [\omega^2 +(2\pi T_L)^2 l^2]\nn
&&~~~~~~\times [\omega^2 +(2\pi T_R)^2 1^2][\omega^2 +(2\pi T_R)^2 3^2]\dots [\omega^2 +(2\pi T_R)^2 l^2]\nn
&&~~~~~~\times {1\over (e^{\omega\over 2 T_L}+1)(e^{\omega\over 2 T_L}+1)} \label{Eqn:GravGammaOddl}
\eea
and for $l$ even 
\bea
d \Gamma_l&=& {\pi\over 8}{(Q_1Q_5)^{l+1}\over 2^{4l} [l!(l+1)!]^2}
\omega^{2l+4}d\omega\nn
&&~~~~~~\times [\omega^2 +(2\pi T_L)^2 2^2] [\omega^2 +(2\pi T_L)^2 4^2]\dots [\omega^2 +(2\pi T_L)^2 l^2]\nn
&&~~~~~~\times [\omega^2 +(2\pi T_R)^2 2^2] [\omega^2 +(2\pi T_R)^2 4^2]\dots [\omega^2 +(2\pi T_R)^2 l^2]\nn
&&~~~~~~\times {1\over (e^{\omega\over 2 T_L}-1)(e^{\omega\over 2 T_L}-1)} \label{Eqn:GravGammaEvenl}
\eea

\subsection{Emission from the CFT state}

In section \ref{vertex} we described the interaction which leads to the emission of a scalar from the CFT state. For a  process where given CFT excitations lead to the emission of this scalar the amplitude per unit time ${\cal R}$ was given in (\ref{rr}).

\subsubsection{The Fermi golden rule for emission}

Let us start with the state where we have no scalar quanta at $t=0$. The amplitude per unit time for the CFT excitations to produce the scalar mode is ${\cal R}$. Let the scalar have energy $\omega_0$, and let the net energy lost by the CFT excitations be $\omega$. After a time $T$ the amplitude for the scalar mode to be excited is
\be
{\cal A}=-i{\cal R} \int _{t=0}^T dt e^{-i\omega t} e^{-i\omega_0 (T-t)}=
-2i{\cal R} e^{-i\omega_0 T}e^{i\Delta\omega T/2}{\sin({\Delta\omega T\over 2})\over \Delta\omega}
\ee
where $\Delta\omega=\omega_0-\omega$. The probability for having emitted the scalar at time $T$ is then
\be
P_{\omega_0}=4|{\cal R}|^2 {\sin^2({\Delta\omega T\over 2})\over (\Delta \omega)^2}
\ee
We can now sum over the closely spaced levels of the emitted scalar (this spacing goes to zero as we take the size of our spherical box $r_{max}$ to infinity). This sum is, from (\ref{ppthtwo})
\be
\sum_{\omega_0} \rightarrow {r_{max}\over \pi}\int d\omega_0
\ee
At large $T$ we have
\be
P_{\omega_0} \rightarrow |{\cal R}|^22\pi T\delta(\omega-\omega_0)
\label{deltaf}
\ee
The sum over scalar levels then gives for the total probability of emission after time $T$ from the particular CFT excitation modes:
\be
P=  2\pi T {r_{max}\over \pi}  \int  d\omega_0 [|{\cal R}|^2\delta(\omega-\omega_0)]=
2\pi T {r_{max}\over \pi} |{\cal R}|^2
\ee
Putting in the value of ${\cal R}$ we get
\be
P= T\f{|V(l)|^2}{8 \pi} \f{\omega^{2(l+1)} \omega_1 \bar{\omega}_1}{(2 \pi R)^{2l+1}} ~  {\delta}_{\sum n_i - \sum \bar n_i - \lambda,0} \prod \mathcal{D}^2
\ee 
To get the emission rate from the CFT state we divide by $T$
\be
\Gamma_{state}=\f{|V(l)|^2}{8 \pi} \f{\omega^{2l+2} \omega_1 \bar{\omega}_1}{(2 \pi R)^{2l+1}} ~  \delta_{\sum n_i - \sum \bar n_i - \lambda,0} \prod \mathcal{D}^2
\label{gammafinal}
\ee

\subsubsection{The sum over states}

The energy levels of the CFT are discrete. Thus the $\delta$-function in  (\ref{deltaf}) will give only discrete values for the energy $\omega$ of the emitted scalar. But for a given scalar energy $\omega_0$ there can be many processes in the CFT which will emit precisely this energy. We must sum over the radiation rates from these processes to find the rate of radiation of scalars with energy $\omega_0$.

We will assume that the scalar has no momentum along the $S^1$ direction; thus $\lambda=0$. Then the $\delta$-function in ${\cal R}$ tells us that $\sum n_i=\sum \bar n_i$, or in terms of energies
\be
\sum_i \omega_i=\sum_i \bar \omega_i
\ee
Then the energy $\delta$-function  in (\ref{deltaf}) tells us that for the process that will contribute to this emission we will have
\be
\sum_i\omega_i=\sum_i \bar \omega_i={\omega \over 2}
\ee
Thus for emission of scalars with one of the allowed discrete energies, we have to compute
\be
\Gamma(\omega)= \sum_{states} \Gamma_{state}=  \f{|V(l)|^2}{8 \pi} \f{\omega^{2l+2} }{(2 \pi R)^{2l+1}}  \sum_{states} \delta_{\sum\omega_i, {\omega \over 2}}~\delta_{\sum \bar \omega_i, {\omega \over 2}}~\omega_1 \bar{\omega}_1~\prod \mathcal{D}^2 \label{Eqn:GammaOmega}
\ee

In section \ref{qwe} we will apply this relation to the emission from a generic state of the near-extremal D1-D5 black hole, thus determining $V(l)$. In section \ref{qwer} we will apply the same relation to the CFT microstates dual to the geometries of \cite{ross}.

\subsection{Emission from generic states: Hawking radiation}\label{qwe}

Let us first compute the emission from the CFT when the CFT state is a generic one for its given total energy. In this case it is known that we will get thermal-looking radiation, which will agree in its coarse-grained properties with the Hawking radiation from the corresponding black hole. 
In this computation the lengths of the component strings are not important to leading order. We have a `thermal gas' of excitations on the component strings, with this `gas' living in a box with length given by the total length of the component string. But the box size does not affect the properties of the gas to leading order, as long as the wavelength of the typical excitation is much less than the length of the box. For simplicity
let us assume that all component strings are singly wound; it can be seen that changing this assumption will not affect the result that we obtain below.

If the component strings are all untwisted,  then the energy levels are spaced ${1\over R}$ apart, and we can replace sums over excitation energies by integrals according to
\be
\sum_\omega\rightarrow R\int d\omega
\ee
In this limit we get from \bref{Eqn:GammaOmega}
\be
 \Gamma(\omega) =  \f{|V(l)|^2}{8 \pi} \f{\omega^{2l+2} }{(2 \pi)^{2l+1} R} \left[\int_\infty^\infty \prod_{i=1}^{l+1}d\omega_i  d \bar \omega_i  ~ \omega_1 \bar \omega_1~ \delta({\omega\over 2}-\sum \omega_i) \delta({\omega\over 2}-\sum \bar \omega_i)  ~ \prod \mathcal{D}^2 \right] \label{Eqn:GammaOmegaContinuous}
 \ee

\subsubsection{The distribution functions $\mathcal D$}

Let us compute the distributions ${\mathcal D}$ for this thermal distribution of excitations. The definitions of these distributions were given at the end of section \ref{vertex}. We will use a method noted in \cite{callan} to simplify their expressions. 

The excitations in the CFT are given by thermal bose and fermi distributions, with the left movers having temperature $T_L$ and the right movers having temperature $T_R$. First consider the left boson $X$. If we annihilate this boson, then for its energy we have $\omega_1>0$, and 

\be
|\mathcal D|^2=<n>=\rho_B(\omega_1)={1\over e^{\omega_1\over T_L}-1}
\ee
Note that the vertex (\ref{vertex}) had a factor $\omega_1$. Summing over the possible $\omega_1$ we will thus get an integral
\be
\int_0^\infty d\omega_1 \omega_1 {1\over e^{\omega_1\over T_L}-1}
\label{ppthir}
\ee
If we create the boson, then we have an energy  $\omega_1<0$ and
\be
|\mathcal D|^2=<n'+1>=\rho_B(-\omega_1)+1={1\over 1-e^{{\omega_1\over T_L}}}
\ee
and summing over the possible $\bar \omega_1$ we will get an integral
\be
\int_{-\infty}^0 d\omega_1 \omega_1 {1\over e^{\omega_1\over T_L}-1}
\ee
We can combine the two integrals above to get
\be
\int_{-\infty}^\infty d\omega_1 \omega_1 {1\over e^{\omega_1\over T_L}-1}=\int_{-\infty}^\infty d\omega_1 \omega_1\rho_B(\omega_1)
\label{ppfourt}
\ee

Similarly, suppose that a fermion is annihilated in the initial state. The energy of the fermion is $\omega_2>0$. We will get  $|\mathcal D|^2=n$, where $n$ is the occupation number of the fermion mode. This will result in an integral
\be
\int_0^\infty d\omega_2 \rho_F(\omega_2)=\int_0^\infty d\omega_2 {1\over e^{\omega_2\over T_L}+1}
\label{ppsixt}
\ee
Creating the fermion will correspond to an energy $\omega_2<0$, and will give a factor $|\mathcal D|^2=1-n$. This gives an integral
\be
\int_0^\infty d\omega_2 (1-\rho_F(-\omega'_2))=\int_{-\infty}^0 d\omega_2 {1\over 1+e^{{\omega'_2\over T_L}}}
\label{ppsevt}
\ee
Combining with (\ref{ppsixt}) we get
\be
\int_{-\infty}^\infty d\omega_2 {1\over e^{\omega_2\over T_L}-1}=\int_{-\infty}^\infty d\omega_2\rho_F(\omega_2)
\ee

\subsubsection{Finding $V(l)$ from Hawking radiation}

Using the distribution functions from the previous section and \bref{Eqn:GammaOmegaContinuous} we get

\bea
\Gamma(\omega) = \f{|V(l)|^2}{8 \pi} \f{\omega^{2l+2} }{(2 \pi)^{2l+1}R}  &\times& [\int_\infty^\infty \prod_{i=1}^{l+1}d\omega_i \rho_B(\omega_1) \prod_{j=2}^{l+1} \rho_F(\omega_j)\omega_1\delta({\omega\over 2}-\sum \omega_i)] \nonumber \\
 &\times&
[\int_\infty^\infty \prod_{i=1}^{l+1}d\bar{\omega}_i \bar{\omega}_1 \rho_B(\bar{\omega}_1) \prod_{j=2}^{l+1} \rho_F(\bar{\omega}_j) \delta({\omega\over 2}-\sum \bar \omega_i)]  \label{Eqn:GammaOmegaThree}
\eea
It can be shown that
\bea
&&\int \prod_{i=1}^{l+1}d\omega_i \rho_B(\omega_1) \prod_{j=2}^{l+1} \rho_F(\omega_j)\omega_1\delta({\omega\over 2}-\sum \omega_i) \nonumber \\
&&\qquad \qquad \qquad = {1\over (l+1)! 2^{l+1}}{1 \over e^{{\omega\over 2T_L}}+1}
 [\omega^2 +(2\pi T_L)^2 1^2][\omega^2 +(2\pi T_L)^2 2^2]\dots [\omega^2 +(2\pi T_L)^2 l^2] \nonumber \\ 
 && \qquad \qquad \qquad \qquad \qquad \qquad \qquad \qquad \qquad \qquad \qquad \qquad  \qquad \qquad \qquad \qquad  \qquad \qquad \text{odd $l$} \nonumber \\
&&\qquad \qquad \qquad =
{1\over (l+1)! 2^{l+1}}{\omega \over e^{{\omega\over 2T_L}}-1}
 [\omega^2 +(2\pi T_L)^2 2^2][\omega^2 +(2\pi T_L)^2 4^2]\dots [\omega^2 +(2\pi T_L)^2 l^2] \nonumber \\
  && \qquad \qquad \qquad \qquad \qquad \qquad \qquad \qquad \qquad \qquad \qquad \qquad  \qquad \qquad \qquad \qquad  \qquad \qquad \text{even $l$} \nonumber \\
 \eea
 Thus from \bref{Eqn:GammaOmegaThree} we get for odd $l$
 \bea
\Gamma(\omega)&=&\f{ |V(l)|^2 \omega^{2l+2}}{8 \pi (2 \pi )^{2l+1}R} \f{1}{[(l+1)!]^2 2^{2(l+1)}}\nn
&&~~~~~~\times [\omega^2 +(2\pi T_L)^2 1^2][\omega^2 +(2\pi T_L)^2 3^2]\dots [\omega^2 +(2\pi T_L)^2 l^2]\nn
&&~~~~~~\times [\omega^2 +(2\pi T_R)^2 1^2][\omega^2 +(2\pi T_R)^2 3^2]\dots [\omega^2 +(2\pi T_R)^2 l^2]\nn
&&~~~~~~\times {1\over (e^{\omega\over 2 T_L}+1)(e^{\omega\over 2 T_L}+1)}
\eea
and for even $l$
\bea
\Gamma(\omega)&=&\f{ |V(l)|^2 \omega^{2l+4}}{8 \pi (2 \pi )^{2l+1}R} \f{1}{[(l+1)!]^2 2^{2(l+1)}}\nn
&&~~~~~~\times [\omega^2 +(2\pi T_L)^2 2^2] [\omega^2 +(2\pi T_L)^2 4^2]\dots [\omega^2 +(2\pi T_L)^2 l^2]\nn
&&~~~~~~\times [\omega^2 +(2\pi T_R)^2 2^2] [\omega^2 +(2\pi T_R)^2 4^2]\dots [\omega^2 +(2\pi T_R)^2 l^2]\nn
&&~~~~~~\times {1\over (e^{\omega\over 2 T_L}-1)(e^{\omega\over 2 T_L}-1)}
\eea
The above expression gives the rate of emission into one of the discrete energies $\omega$ allowed for the outgoing scalar. The classical emission rate is however given as a continuous function of the energy. Thus consider a narrow band of width $d\omega$ around $\omega$. Since the left and right excitations on the CFT occur in units of ${1\over R}$ each, the energies of the emitted scalar are in units of ${2\over R}$. Thus if we sum  the rate of emission over all the scalar energies in the range $d\omega$ we will get a factor
\be
\sum_\omega\rightarrow {R\over 2} d\omega
\ee
The rate of emission into the range $d\omega$ for odd l then is
 \bea
d \Gamma_l&=&\f{ |V(l)|^2 \omega^{2l+2} ~d\omega }{8 \pi (2 \pi )^{2l+1}} \f{1}{[(l+1)!]^2 2^{2l+3}}\nn
&&~~~~~~\times [\omega^2 +(2\pi T_L)^2 1^2][\omega^2 +(2\pi T_L)^2 3^2]\dots [\omega^2 +(2\pi T_L)^2 l^2]\nn
&&~~~~~~\times [\omega^2 +(2\pi T_R)^2 1^2][\omega^2 +(2\pi T_R)^2 3^2]\dots [\omega^2 +(2\pi T_R)^2 l^2]\nn
&&~~~~~~\times {1\over (e^{\omega\over 2 T_L}+1)(e^{\omega\over 2 T_L}+1)}
\eea
and for even $l$
\bea
d \Gamma_l&=&\f{ |V(l)|^2 \omega^{2l+4} ~d\omega}{8 \pi (2 \pi )^{2l+1}} \f{1}{[(l+1)!]^2 2^{2l+3}}\nn
&&~~~~~~\times [\omega^2 +(2\pi T_L)^2 2^2] [\omega^2 +(2\pi T_L)^2 4^2]\dots [\omega^2 +(2\pi T_L)^2 l^2]\nn
&&~~~~~~\times [\omega^2 +(2\pi T_R)^2 2^2] [\omega^2 +(2\pi T_R)^2 4^2]\dots [\omega^2 +(2\pi T_R)^2 l^2]\nn
&&~~~~~~\times {1\over (e^{\omega\over 2 T_L}-1)(e^{\omega\over 2 T_L}-1)}
\eea
We can now compare this CFT emission rate to the classical emission rate (\ref{Eqn:GravGammaOddl},\ref{Eqn:GravGammaEvenl}). This comparison fixes the normalization of the CFT interaction vertex:
\be
 |V(l)|^2
 ={16 \pi^{2l+3}  \over [l!]^2} (Q_1Q_5)^{l+1}
 \label{normalization}
\ee

\subsection{Emission from the special microstate}\label{qwer}

Having fixed the normalization of the vertex (\ref{normalization}), we can now look at the emission process (\ref{omegaimq}) from the special microstates of \cite{ross}. Consider the expression for the CFT emission rate $\Gamma$ given in (\ref{gammafinal}). There is only one particular CFT process that will lead to the required emission. We will create the bosons $X, \bar X$, and there are no bosons in the CFT state before the emission. Thus we have $\mathcal D(\omega_1)=\mathcal D(\bar\omega_1)=1$.  The fermions in the initial state are annihilated, from levels where they were known to be present; thus for the fermions we also get $\mathcal D (\omega_i)=\mathcal D(\bar\omega_i)=1$.   As we saw in section \ref{ob}, the created bosons bosons will be  in the lowest allowed energy level, which is
\be
\omega_1=\bar\omega_1={1\over R}
\ee
Then we find that
\be
\Gamma=\f{ |V(l)|^2 \omega^{2l+2}}{8 \pi (2 \pi R)^{2l+1}} {1\over R^2}=\f{4 \pi}{[l!]^2} \left ( \f{Q_1Q_5}{4 R^2} \omega^2 \right)^{l+1} \f{1}{R} 
\ee
Comparing with (\ref{omegaimq}) we then observe that
\be
\Gamma=2\omega_I
\ee
This establishes the required relation between radiation from the CFT state and the energy radiated by the classical instability. It should be mentioned however that this computation has been a somewhat heuristic one, since we have used the  heuristic vertex operator construction given in section \ref{vertex}. As mentioned before, one should really perform a rigorous construction of the emission vertex along the lines of the constructions in \cite{lm2,lmpre};  an exercise that we hope to return to elsewhere.\footnote{For example, the heuristic vertex construction does not address the `base spin' of the component strings in an adequate fashion. When the component strings are twisted together, the base spin gets changed, and leads to the creation of fermions in the lowest levels of the fermi sea. The heuristic vertex assumes that the amplitude for this operation is the same for all transitions.}

To summarize, the CFT computation shows that the classical instability of \cite{myers} is just the emission expected from the microstates of \cite{ross}, if we consider  exactly the same process that leads to Hawking emission from {\it  generic} states of the black hole.

\section{Discussion}\label{disc}
\setcounter{equation}{0}

Let us discuss the implications of our results for the fuzzball proposal of black holes.

\subsection{Extremal holes}

\begin{figure}[htbp] 
   \centering
   \includegraphics[width=4.7in]{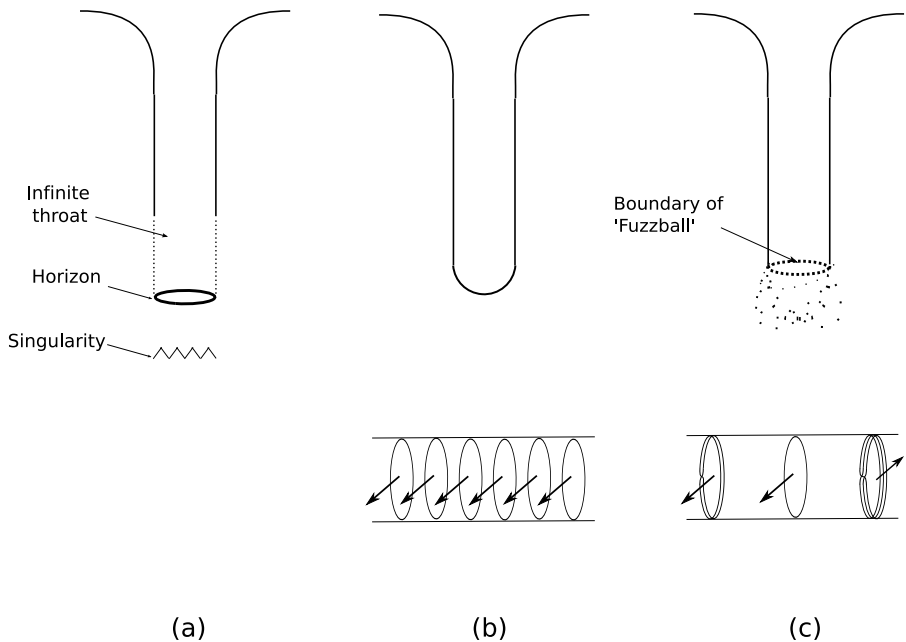} 
   \caption{(a) Traditional picture of the extremal hole (b) Geometry of a special microstate; all CFT component strings (shown in the lower diagram) are are in the same state, and this makes the geometry classical (c) A generic microstate; the CFT component strings are have a distributions of lengths and spins.}
   \label{Fig:ex}
\end{figure}

First let us recall the fuzzball picture for the case of extremal holes. In fig.\ref{Fig:ex}(a) we depict the conventional picture of the hole. We have flat space at infinity, then an infinite throat; the throat ends in a horizon, and there is a singularity inside the horizon. The important point here is that there is no information about the state of the hole in the vicinity of the horizon: we just have vacuum spacetime there. It can be argued that there is no `energy gap' in this system: we can place a quantum deep down the infinite throat and make an excitation of the hole with as low an energy as we wish. In this picture we have information loss: if we throw in a quantum then its information does not return in the Hawking radiation
it creates. It is also strange that an object (the black hole) with finite mass and located in a finite region of space exhibits no `energy gap'; one would expect a small but nonzero gap for any such object.

Several families of microstates of extremal holes have been constructed, and they  do {\it not} agree with the picture in fig.\ref{Fig:ex}(a). In fig.\ref{Fig:ex}(b) we depict a particular (nongeneric) microstate of the hole. In the CFT dual we have all component strings of the same length and with the same spin. Choosing all these component strings to be in the same state makes the corresponding geometry classical. Since the throat is now `capped off', there is a finite energy gap. This gap is found to agree {\it exactly} with the gap in the dual CFT state \cite{gms2}.

In fig.\ref{Fig:ex}(c) we depict the generic microstate. The throat is long but not infinite, and ends in a quantum `fuzz'. The CFT dual has component strings with a wide variety of lengths and spins; this makes the corresponding geometry have large quantum fluctuations, and it is not well characterized by any classical metric. A test quantum that falls past the dotted circle stays trapped in the fuzz for long times, and so for an observer working on small classical time scales it appears that the quantum has fallen past a `horizon' and cannot return. But since the information eventually returns  from the `fuzzball' (over the long Hawking evaporation timescale), there is no information loss. The large change from fig.\ref{Fig:ex}(a) to \ref{Fig:ex}(c) is possible because in string theory we have the phenomenon of `fractionation', which generates very low energy `fractional brane excitations'  when a large number of quanta are bound together \cite{emission}. 

\subsection{Non-extremal holes}

\begin{figure}[htbp] 
   \centering
   \includegraphics[width=4.7in]{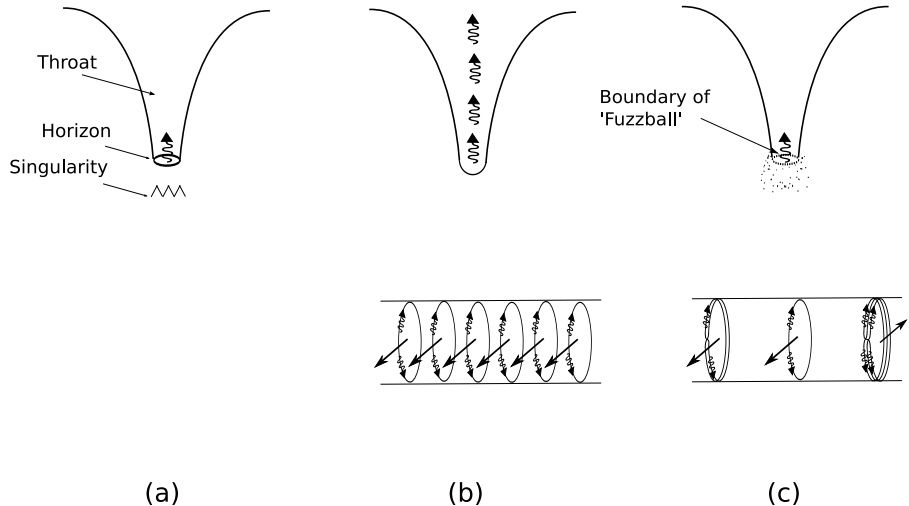} 
   \caption{(a) Traditional picture of the non-extremal hole; Hawking radiation emerges from the horizon where there is  no information of the black hole state (b) Radiation from the microstates of \cite{ross}; the non-extremal energy is emitted through a classical instability. The CFT state has all component strings in the same excited state, and Bose enhancement leads to a classical radiation rate (c) Our expectation for the generic non-extremal microstate. The CFT component strings have a distribution of  lengths, spins and excitations, and thermal looking radiation emerges slowly. In the gravity description the complicated `cap' region leaks the non-extremal energy slowly as information-carrying Hawking radiation. 
}
   \label{Fig:non}
\end{figure}

In fig.\ref{Fig:non}(a) we depict the traditional picture of the non-extremal hole. The throat is not infinite; it ends in a horizon after a finite distance, and there is a singularity inside the horizon.   In fig.\ref{Fig:non}(b) we depict a nongeneric microstate geometry of the kind made in \cite{ross}. In the CFT dual all component strings have the same length and carry the same excitations. The fact that we have  a large number of component strings of the same kind makes the dual geometry classical. It also makes the radiation from the state very strong because of Bose enhancement, as discussed in section \ref{radiationrate}. We thus see a classical instability.

In fig.\ref{Fig:non}(c) we depict our expectation for the generic microstate for the non-extremal hole. The component strings have a wide dispersion in their  lengths and excitations, so there is no Bose enhancement. The {\it same} computation that produces the instability in fig.\ref{Fig:non}(b) now gives a planckian radiation spectrum.\footnote{Some consequences of this spread in lengths of component strings were discussed in \cite{mathuressay}.}  This radiation from the generic CFT state is known to reproduce all details of Hawking radiation from the corresponding black hole, like spin dependence and greybody factors \cite{callanmalda,dasmathur,maldastrom}. We now see how this radiation would emerge in the gravity picture: it leaks out slowly from the complicated, quantum `cap' region of the geometry, just as the energy radiated out of the `cap' region in fig.\ref{Fig:non}(b); the only difference is that the simple `cap' of fig.\ref{Fig:non}(b) allowed a much more rapid escape of energy.

\bigskip

The computation of this paper advances the fuzzball proposal in several ways.  First, since we are working with non-extremal geometries, we can see radiation. Thus we have an explicit microstate geometry (constructed in \cite{ross}) and we see the appropriate `Hawking radiation' from this microstate. In earlier work with extremal microstates we could only study small departures from extremality by throwing test quanta into the geometry.

Secondly, we see that the microstate geometries we have used should be thought of as non-generic members of the ensemble of black hole geometries, rather than as special states unrelated to black holes. To emphasize this point we  started with the semiclassical computation of Hawking radiation from the black hole, and used it to find the interaction strength $V(l)$ in the CFT.  We then used this $V(l)$ to find the radiation from our non-generic state, and reproduced the classical instability seen there.
Thus in the CFT picture we can track explicitly the changes in the behavior of the state as we replace generic distributions of component strings with special distributions. 

Lastly, we have seen an interesting phenomenon of Bose enhancement in the radiation process. We saw that once we had created $n$ twisted component strings in our system, the probability to create the next one was proportional to $(n+1)$; this gave rise to the exponential growth of the perturbation in our chosen microstates. It would be interesting to investigate how this phenomenon appears when we start with more generic microstates; in particular whether there is always an evolution towards having many identical component strings in the system because of Bose enhancement. One may also be able to use AdS/CFT duality with different microstate geometries to learn about properties of Bose condensates, just as the AdS-Schwarzschild black hole geometry is used to analyze properties of thermal QCD. We hope to return to these issues elsewhere.

\section*{Acknowledgements}
\setcounter{equation}{0}

We thank Steve Avery, Eric Gimon, Stefano Giusto, Jeremy Michelson, Mohit Randeria, Simon Ross,  Kostas Skenderis, Rakesh Tiwari and Nandini Trivedi  for many helpful comments. 
This work was supported in part by DOE grant DE-FG02-91ER-40690.

\appendix
\section{Solving the wave equation by `matching'}
\label{one}
\renewcommand{\theequation}{A.\arabic{equation}}
\setcounter{equation}{0}

The radial equation is (\ref{radial})
\be
\partial_x[ x(x+1) \partial_x h] + \f{1}{4} \left[ \kappa^2 x + 1-\nu^2 + \f{\xi^2}{x+1} -\f{\zeta^2}{x} \right]h=0
\ee

\subsection{Inner Region} 
In the inner region the equation is
\be
4 \partial_x[ x(x+1) \partial_x h] + \left[  1-\nu^2 + \f{\xi^2}{x+1} -\f{\zeta^2}{x} \right]h=0
\ee
Writing 
\be
h=x^\f{\zeta}{2} (1+x)^\f{\xi}{2} w
\ee
we get 
\be
x(1+x) \partial_x^2 w + (c+(a+b+1)x) \partial_x w + a b w=0
\ee
with
\be
a=\h(1+ \zeta + \xi + \nu), \qquad b=\h(1+ \zeta + \xi -\nu), \qquad c=1+ \zeta
\ee
The solution is
\bea
h&=& x^\f{\zeta}{2} (1+x)^\f{\xi}{2} \left[ D_1 \p_2 F_1(a,b,c,-x) +  D_2  \p_2 x^{1-c} F_1(1+a-c,1+b-c,2-c,-x) \right]  \\
&=&  (1+x)^\f{\xi}{2} \left[ x^\f{\zeta}{2} D_1 \p_2 F_1(a,b,c,-x) + D_2 x^\f{-\zeta}{2} F_1(1+a-c,1+b-c,2-c,-x)  \right] \nonumber
\eea
Demanding regularity of the solution at $x \to 0$ we get 
\be
h=(1+x)^\f{\xi}{2} x^\f{|\zeta|}{2} D \p_2 F_1(a,b,c,-x) 
\ee

To find the behavior in the matching region we use the following identity for hypergeometric functions
\bea
\p_2F_1(a,b,c,-x) &=& \f{\Gamma(c) \Gamma(b-a)}{\Gamma(b) \Gamma(c-a)} x^{-a}  \p_2 F_1 (a,1-c+a,1-b+a,-\f{1}{x}) \\
&+& \f{\Gamma(c) \Gamma(a-b)}{\Gamma(a) \Gamma(c-b)} x^{-b}  \p_2 F_1 (b,1-c+b,1-a+b,-\f{1}{x}) \nonumber
\eea
This gives
\bea
h &=& (1+x)^\f{\xi}{2} x^\f{|\zeta|}{2} D \left[ \f{\Gamma(c) \Gamma(b-a)}{\Gamma(b) \Gamma(c-a)} x^{-\h(1+|\zeta|+\xi + \nu)} +  \f{\Gamma(c) \Gamma(a-b)}{\Gamma(a) \Gamma(c-b)} x^{-\h(1+|\zeta|+\xi - \nu)} \right] \\
&=& D \Gamma(1+ |\zeta|) \Big[ \f{\Gamma(-\nu)}{\Gamma(\h(1+ |\zeta| +\xi -\nu))\Gamma(\h(1+ |\zeta| -\xi -\nu))} x^{-\f{\nu+1}{2}} \nonumber \\
&+&\f{\Gamma(\nu)}{\Gamma(\h(1+ |\zeta| +\xi +\nu))\Gamma(\h(1+ |\zeta| -\xi +\nu))} x^{\f{\nu-1}{2}}  \Big] \nonumber
\eea

\subsection { Outer Region} 
The equation in the outer region is
\be
4 \partial_x[ x^2 \partial_x h]  + \kappa^2 x h + (1-\nu^2) h=0
\ee
Writing $z= \sqrt{x}$ and $h = z w$ we get
\be
\partial_{\kappa z}^2 w + \f{1}{ \kappa z} \partial_{\kappa z} w + (1 - \f{\nu^2}{\kappa^2 z^2}) w =0
\ee
This has the solution
\be
h = \f{1}{\sqrt{x}} \left[ C_1 J_\nu (\kappa \sqrt{x}) + C_2 J_{-\nu} (\kappa \sqrt{x}) \right]
\ee
At large $x$  we get 
\be
h \sim \f{1}{x^\f{3}{4}} \f{1}{\sqrt{2 \pi  \kappa}} [ e^{i \kappa \sqrt{x} }e^{-i \f{\pi}{4}}( C_1 e^{-i \nu \f{\pi}{2}} + C_2 e^{i \nu \f{\pi}{2}} )+e^{-i \kappa \sqrt{x}} e^{i \f{\pi}{4}}( C_1 e^{i \nu \f{\pi}{2}} + C_2 e^{-i \nu \f{\pi}{2}} ) ] \label{Eqn:OutOuterSol}
\ee
and at small $x$
\be
h \sim  C_1  \f{1}{\Gamma(1+ \nu)} \left( \f{\kappa}{2} \right)^\nu x^\f{\nu-1}{2}+C_2  \f{1}{\Gamma(1- \nu)} \left( \f{\kappa}{2} \right)^{-\nu} x^{-\f{\nu+1}{2}}
\ee

\subsection {Matching the solutions} 

Matching the  solutions from the outer and inner regions  we get
\be
\f{C_1}{C_2} \f{\Gamma(1- \nu)}{\Gamma(1+ \nu)} \left(\f{\kappa}{2} \right)^{2\nu} = \f{\Gamma(\nu)}{\Gamma(-\nu)} \f{\Gamma(\h(1+ |\zeta| +\xi -\nu))\Gamma(\h(1+ |\zeta| -\xi -\nu)) }{   \Gamma(\h(1+ |\zeta| +\xi +\nu))\Gamma(\h(1+ |\zeta| -\xi +\nu))   }
\ee

\subsection{ The instability}

The geometry has an instability if we have no incoming wave, but yet we have an outgoing wave carrying energy out to infinity.   We see from \bref{Eqn:OutOuterSol} that the if we have no incoming wave we get the relation

\be
C_1 + C_2 e^{- i \pi \nu} = 0
\ee
Thus the frequencies of the instabilities  are given by solutions to the equation
\be
-e^{-i \nu \pi} \f{\Gamma(1- \nu)}{\Gamma(1+ \nu)} \left(\f{\kappa}{2} \right)^{2\nu} = \f{\Gamma(\nu)}{\Gamma(-\nu)} \f{\Gamma(\h(1+ |\zeta| +\xi -\nu))\Gamma(\h(1+ |\zeta| -\xi -\nu)) }{   \Gamma(\h(1+ |\zeta| +\xi +\nu))\Gamma(\h(1+ |\zeta| -\xi +\nu))   }
\label{basic}
\ee

\section{Obtaining the instability frequencies}
\label{two}
\renewcommand{\theequation}{B.\arabic{equation}}
\setcounter{equation}{0}

We will analyze (\ref{basic}) in the limit of large $R$ which is the limit where we have a CFT description of the physics. In this limit, using  (\ref{kappasq}) and (\ref{Eqn:LargeRDiffHorizon}), we will have
\be
\kappa^2\sim (\omega^2-\f{\lambda^2}{R^2}){Q_1Q_5\over R^2} <<1 \label{Eqn:smallKappa}
\ee
We  follow the method used in \cite{myers}  for solving (\ref{basic}) in this limit. One first observes that the LHS of (\ref{basic}) is small because $\kappa^2$ is small. The RHS can be small if one of the $\Gamma$ functions in the denominator is close to having a pole. At the first order of iteration we will set the argument of this $\gamma$ function so that we do get a pole; at the next order of iteration we see how far we need to be off this pole to get the small but nonvanishing LHS.

We let the term $\Gamma(\h(1+ |\zeta| +\xi+\nu)) $ have a pole by setting
\be
1+ |\zeta| +\xi +\nu = -2 N \label{Eqn:pole}
\ee
with $N \ge 0$ an integer. Let us recall  \bref{Eqn:Definitions1} where the terms in this equation were defined. At leading order we take $R \to \infty$. The terms in \bref{Eqn:pole} involve $\varrho,\vartheta$ so we first analyze the leading order behavior of these variables. From \bref{Eqn:Definitions1} and \bref{Eqn:LargeRCoshSinhSame} and we see that
\bea
\varrho &\to& 1 \\ 
\vartheta &\to& 0 \nonumber 
\eea
With this we see that at leading order
\bea
\zeta &\simeq & -\lambda -m_\psi n + m_\phi m \\
\xi &\simeq & \omega R  - m_\phi n + m_\psi m \nonumber \\
\nu &\simeq& l+1
\eea
So the the equation \bref{Eqn:pole} becomes
\be
1+ |-\lambda -m_\psi n + m_\phi m| + \omega R  - m_\phi n + m_\psi m + l+1 = -2 N
\ee
This gives for the instability frequencies 
\be
\omega= \f{1}{R}\left(-l   - m_\psi m + m_\phi n - |-\lambda -m_\psi n + m_\phi m| -2(N+1) \right)
\ee
Thus the leading order frequencies are real. To find the imaginary part of $\omega$, we need to iterate to the next order. We change the argument of the divergent $\Gamma$ from an integer $-N$ to $-N-\delta N$. We use \bref{Eqn:pole} to eliminate  $\xi$ in favor of $N$. Then the matching condition becomes
\be
e^{- i \pi \nu } \left(\f{\Gamma(-\nu)}{\Gamma(\nu)} \right)^2 \left(\f{\kappa}{2} \right)^{2\nu} = \f{ \Gamma(-N-\nu - \delta N) \Gamma(N+ 1+ |\zeta| + \delta N)}{\Gamma(-N - \delta N) \Gamma(N+1 + |\zeta| + \nu + \delta N)}
\ee
In the $\delta N \to 0$ limit only $\Gamma(-N-\delta N)$ develops a pole. We can neglect the $\delta N$ correction in the other $\Gamma$ functions to get
\be
e^{- i \pi \nu } \left(\f{\Gamma(-\nu)}{\Gamma(\nu)} \right)^2 \left(\f{\kappa}{2} \right)^{2\nu} = \f{ \Gamma(-N-\nu ) \Gamma(N+ 1+ |\zeta| )}{\Gamma(N+1 + |\zeta| + \nu )} (-1)^{N+1}  N!  \delta N
\ee
where we have used $\Gamma(-n+ \delta n ) = [(-1)^n n! \delta n]^{-1}$ for $n$ a non-negative integer and $\delta n$ a small number. We want to simplify this expression and solve for $\delta N$. We use the relation
\bea
\Gamma(n+1+x) = x n! [x]_n \Gamma(x), \qquad  [x]_n \equiv \prod_{i=1}^n (1+ \f{x}{i})
\eea
where $n$ is a positive integer, to write
\be
\f{\Gamma(N+1+ |\zeta| + \nu)}{\Gamma(N+1+ |\zeta|)} = \nu [\nu]_{N+ |\zeta|} \Gamma(\nu)
\ee
We further use the relation
\be
\Gamma(-n-x) = \f{\Gamma(-x)}{(-1)^n n! [x]_n}
\ee
All this gives
\be
e^{- i \pi \nu } \left(\f{\Gamma(-\nu)}{\Gamma(\nu)} \right)^2 \left(\f{\kappa}{2} \right)^{2\nu} = \f{1}{\nu \Gamma(\nu) [\nu]_{N+|\zeta|} } \f{\Gamma(-\nu)}{(-1)^N N! [\nu]_N} (-1)^{N+1} N! \delta N
\ee
Thus we find
\be
\delta N = - e^{- i \pi \nu } \left( \nu \f{\Gamma(-\nu)}{\Gamma(\nu)} \right) \left(\f{\kappa}{2} \right)^{2\nu} [\nu]_N [\nu]_{N+ |\zeta|}
\ee
$\delta N$ can have an imaginary part because  $e^{-i\pi \nu}$  has an imaginary part $-i\sin (\pi \nu)$.  We use the relation $ \Gamma(\nu) \Gamma(-\nu) = -\f{\pi}{ \nu \sin (\pi \nu)}$ to get
\be
{\mathcal Im} (\delta N) = - \f{\pi}{\Gamma(\nu)^2} \left(\f{\kappa}{2} \right)^{2\nu} [\nu]_N [\nu]_{N+ |\zeta|}
\ee
At leading order $\nu = l+1$ is an integer and it can be seen that for integers $p,q$
\be
[p]_q= \p^{p+q}C_p
\ee
so we get at leading order 
\be
{\mathcal Im} (\delta N) = - \f{\pi}{[l!]^2} \left[(\omega^2-\f{\lambda^2}{R^2}) {Q_1Q_5\over {4 R^2}} \right]^{(l+1)} \p^{l+1+N}C_{l+1} \p^{l+1+ N+ |\zeta|}C_{l+1} \label{Eqn:DeltaImN}
\ee
We now wish to find the relation between $\delta N$ and $\delta \omega$. From  \bref{Eqn:pole} we have
\bea
-2\delta N &=&  \partial_\omega(|\zeta|+  \xi + \nu ) \delta \omega\\
&=&  \partial_\omega( \xi  )\delta \omega \nonumber \\
&=&R ~ \delta \omega 
\eea
where we have used the fact that at leading order only $\xi$ depends on $\omega$. Since at leading order $\omega$ is real, we have
\bea
\omega_I &=& - \f{2 {\mathcal Im}(\delta N)}{R} \nonumber \\
&=& \f{1}{R} \left( \f{2 \pi}{[l!]^2}\left[ (\omega^2-\f{\lambda^2}{R^2})  {Q_1Q_5\over {4 R^2}} \right]^{(l+1)} \p^{l+1+N}C_{l+1} \p^{l+1+ N+ |\zeta|}C_{l+1} \right)
\eea
We see that
\be
\omega_I >0
\ee
which gives an exponential growth of the perturbation.
We arrived at this result by iterating near the poles of $\Gamma(\h(1+ |\zeta| +\xi+\nu)) $. If we iterate near the poles of $\Gamma(\h(1+ |\zeta| -\xi+\nu)) $ we will obtain an $\omega$ with a negative imaginary part, giving an exponentially decaying  perturbation.

\end{document}